\documentclass[]{style/ceurart}
\usepackage{graphicx}
\usepackage{booktabs}
\usepackage{multirow}
\usepackage{subcaption}
\usepackage{listings}
\begin{document}
\copyrightyear{2025}
\copyrightclause{Copyright for this paper by its authors.
  Use permitted under Creative Commons License Attribution 4.0
  International (CC BY 4.0).}

\conference{CLEF 2025 Working Notes, 9 -- 12 September 2025, Madrid, Spain}

\title{Distilling Spectrograms into Tokens: Fast and Lightweight Bioacoustic Classification for BirdCLEF+ 2025}
\title[mode=sub]{Notebook for the LifeCLEF Lab at CLEF 2025}

\author[1]{Anthony Miyaguchi}[
orcid=0000-0002-9165-8718,
email=acmiyaguchi@gatech.edu,
]
\cormark[1]

\author[1]{Murilo Gustineli}[
orcid=0009-0003-9818-496X,
email=murilogustineli@gatech.edu,
url=https://murilogustineli.com,
]
\cormark[1]

\author[1]{Adrian Cheung}[
orcid=0009-0006-8650-4550,
email=acheung@gatech.edu,
]
\cormark[1]

\address[1]{Georgia Institute of Technology, North Ave NW, Atlanta, GA 30332}
\cortext[1]{Corresponding author.}

\begin{abstract}
The BirdCLEF+ 2025 challenge requires classifying 206 species, including birds, mammals, insects, and amphibians, from soundscape recordings under a strict 90-minute CPU-only inference deadline, making many state-of-the-art deep learning approaches impractical.
To address this constraint, the DS@GT BirdCLEF team explored two strategies.
First, we establish competitive baselines by optimizing pre-trained models from the Bioacoustics Model Zoo for CPU inference.
Using TFLite, we achieved a nearly 10x inference speedup for the Perch model, enabling it to run in approximately 16 minutes and achieve a final ROC-AUC score of 0.729 on the public leaderboard post-competition and 0.711 on the private leaderboard.
The best model from the zoo was BirdSetEfficientNetB1, with a public score of 0.810 and a private score of 0.778.
Second, we introduce a novel, lightweight pipeline named Spectrogram Token Skip-Gram (STSG) that treats bioacoustics as a sequence modeling task.
This method converts audio into discrete "spectrogram tokens" by clustering Mel-spectrograms using Faiss K-means and then learns high-quality contextual embeddings for these tokens in an unsupervised manner with a Word2Vec skip-gram model.
For classification, embeddings within a 5-second window are averaged and passed to a linear model.
With a projected inference time of 6 minutes for a 700-minute test set, the STSG approach achieved a final ROC-AUC public score of 0.559 and a private score of 0.520, demonstrating the viability of fast tokenization approaches with static embeddings for bioacoustic classification.
Supporting code for this paper can be found at \url{https://github.com/dsgt-arc/birdclef-2025}.
\end{abstract}

\maketitle
\section{Introduction}

The BirdCLEF+ 2025 challenge in the LifeCLEF Lab involves classifying fauna in the Middle Magdalena Valley of Colombia \cite{birdclef2025} \cite{lifeclef2025}.
Participants are provided several thousand one-minute soundscapes and must predict the probability that one of 206 species appears in non-overlapping 5-second intervals.
While previous iterations of the BirdCLEF challenge have focused on avian calls, this year includes mammals, insects, and amphibians.

One important limitation of solutions is that they must fit within a 90-minute inference deadline on a CPU-only instance provided on Kaggle.
The computational constraint discourages solutions that rely on gross computation to reach the top of the leaderboard through mechanisms like model ensembling.

We address the BirdCLEF+ challenge in two parts.
First, we provide baseline transfer-learning solutions by training a classification head on pre-trained birdcall classification models.
We hypothesize that the representation space of existing bioacoustic models will transfer well to the domain-shifted dataset.
Existing research has shown that birdcall classifiers, such as BirdNET and Perch, handle domain shifts well \cite{ghani2023global} and even transfer to novel environments, including marine environments, with SurfPerch \cite{williams2025using}.
We optimize models for CPU in order to meet competition inference timeouts.

We then experiment with a technique that utilizes discrete audio tokens for soundscape classification, which we refer to as Spectrogram Token Skip-Gram (STSG).
First, we convert audio into discrete tokens derived from Mel-spectrograms. 
We then contextualize tokens into continuous space via skip-gram embeddings and average tokens within prediction intervals to build a classification head, offering some reasonable defaults for the hyperparameters involved.
Finally, we experiment with pretraining a classification model on a surrogate task using a student-teacher model on unlabeled training soundscapes, leveraging predictions from a strong bioacoustical model.
\section{Related Work}

The dominant strategy for bioacoustic classification, particularly in BirdCLEF, involves treating audio as a computer vision problem and applying CNN ensembles to Mel-spectrograms, 2D time-frequency representations of audio signals \cite{birdclef2024}.
Solutions often utilize transfer learning from EfficientNet, ConvNeXt, and similar backbones, as well as specialized bioacoustic models such as BirdNET and Perch \cite{miyaguchi2024transfer}.
However, the increasing computational cost of these large ensembles poses a challenge in the constrained competition setting.
This has made inference optimization a critical component of state-of-the-art methods, particularly in model compilation using TFLite, ONNX, or OpenVINO \cite{birdclef2024}.

In the broader field of audio processing, many newer models represent audio as compact sequences of discrete tokens.
This allows the use of powerful sequence modeling primarily used in the text domain, including LLMs.
These tokens can be divided into two types: \emph{acoustic tokens} and \emph{semantic tokens}.
Neural audio codecs, such as SoundStream \cite{zeghidour2021soundstream} and EnCodec \cite{defossez2022high}, produce acoustic tokens optimized for reconstructing the original waveform, which is often expensive for discriminative tasks.
In contrast, models that produce semantic tokens capture abstract information but require large self-supervised architectures like wav2vec \cite{baevski2020wav2vec}.
While most audio tokenization efforts focus on speech data, this work explores tokenization for resource-constrained scenarios in other audio domains.

The word2vec skip-gram model is a simple method for learning dense vector embeddings from large text corpora \cite{mikolov2013efficient}.
It predicts surrounding context words given a single target word, learning a low-dimensional vector for each word that captures semantic relationships.
Our STSG method adapts this principle to audio sequences, learning to embed spectrogram tokens that frequently co-occur in time close to each other in the embedding space.
This makes downstream classification more efficient, in contrast to more complex models like AudioLM, which use expensive transformers to model sequences of tokens \cite{borsos2023audiolm}.

\section{Methodology}

We split our methods into two main parts.
We explore transfer learning in depth for the BirdCLEF 2025+ dataset using a variety of pre-trained soundscape backbones.
Here, we establish baseline classification and computational performance characteristics on state-of-the-art models and determine the engineering work required to deploy them to resource-constrained systems.
We then explore our central hypothesis that a tokenized representation of spectrograms can be run with marginal classification degradation while reducing inference computation by an order of magnitude.

\begin{figure}
 \centering
 \includegraphics[width=0.8\linewidth]{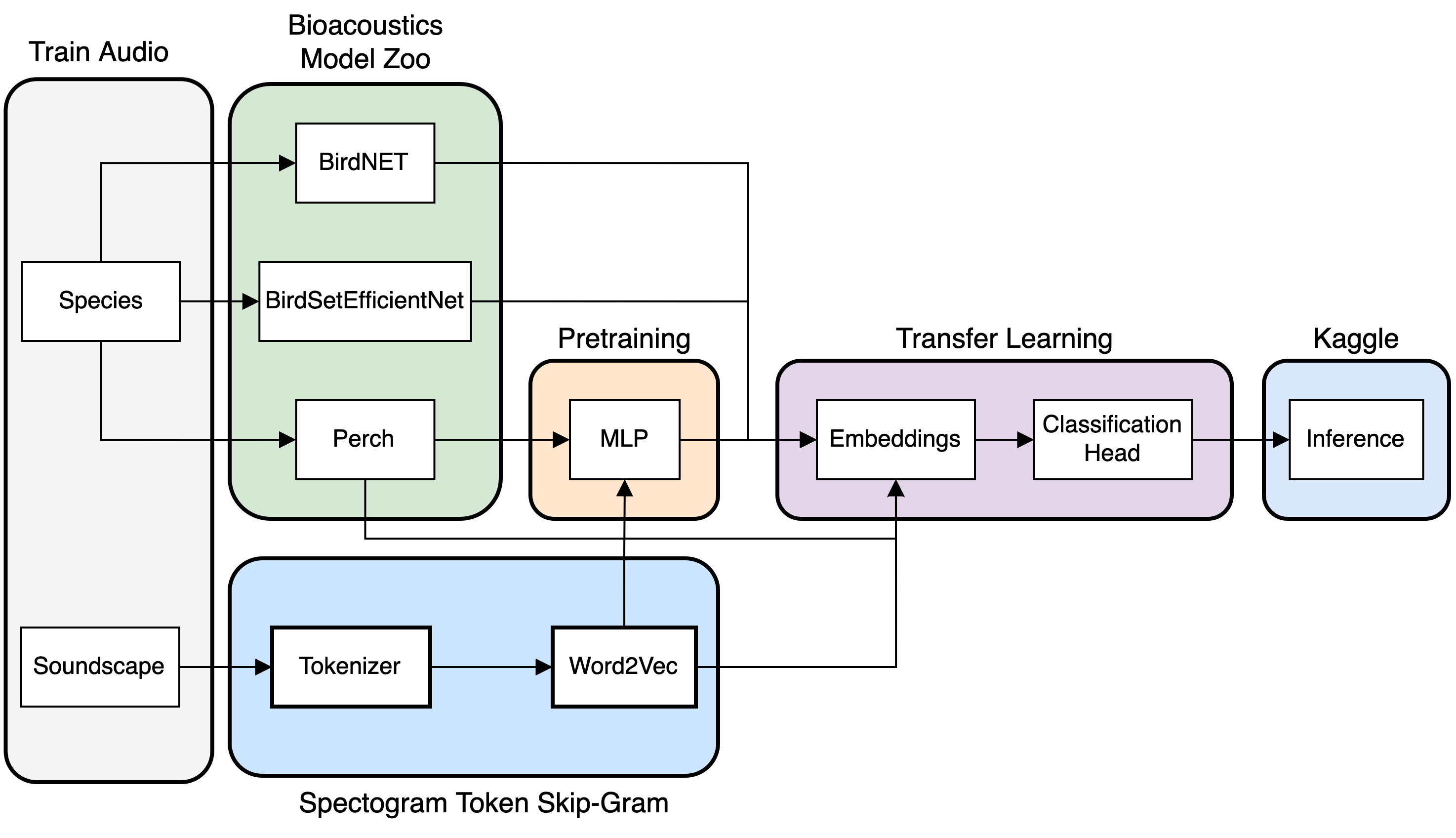}
 \caption{
 Overview of various experiments as a flow diagram.
 We utilize pre-trained bioacoustic classifiers and transfer learning onto the BirdCLEF+ task using a surrogate classification task.
 We also experiment with tokenized MFCC embeddings and student-teacher pre-training on top of these embeddings.
 }
 \label{fig:system-diagram}
\end{figure}

\subsection{Transfer Learning with Bioacoustics Model Zoo}

\begin{figure}[!h]
 \centering
 \includegraphics[width=0.75\columnwidth]{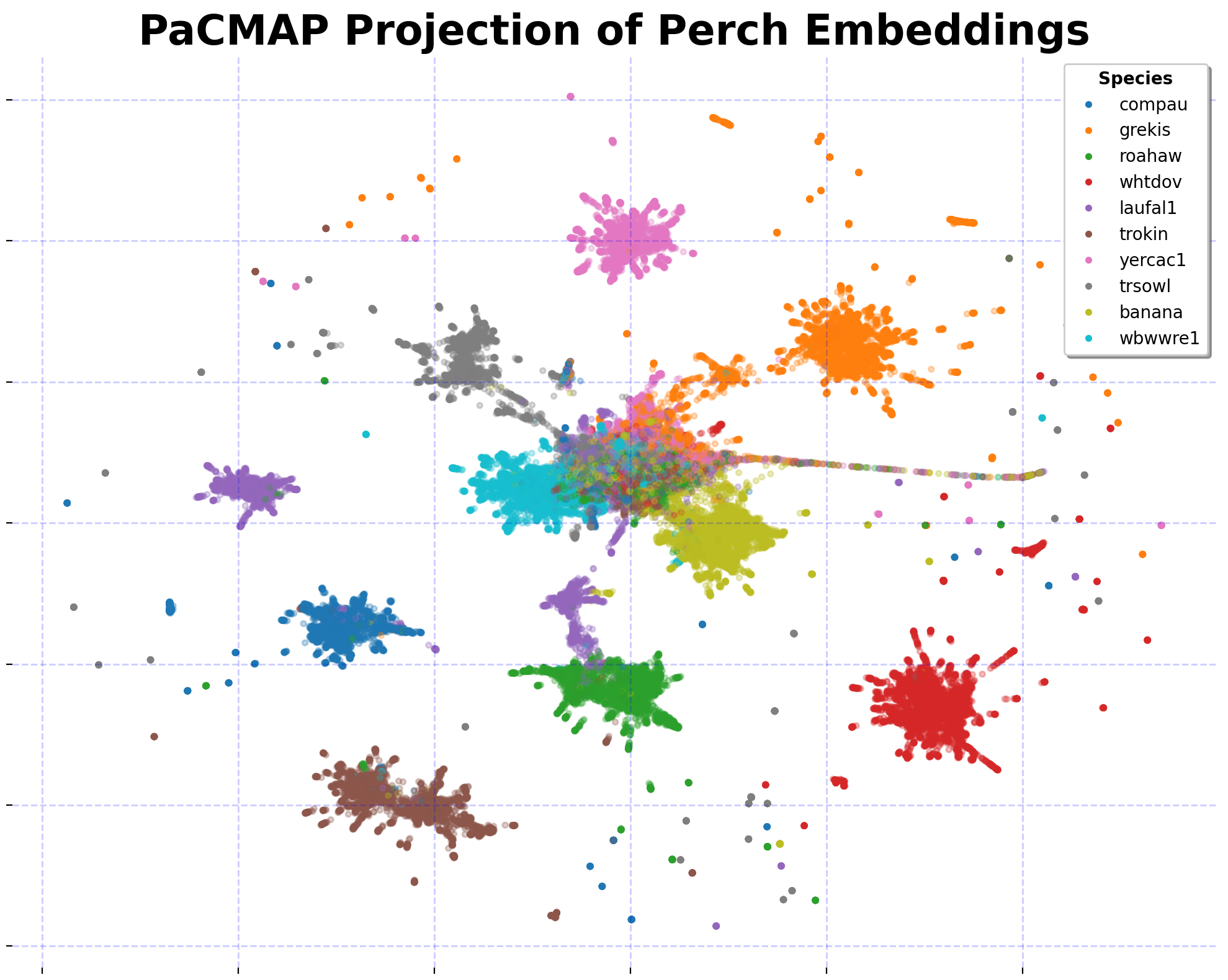}
 \caption{PaCMAP projection showing the distribution of Perch embeddings for the 10 most frequent species: \textbf{\textit{compau}, \textit{grekis}, \textit{roahaw}, \textit{whtdov}, \textit{laufal1}, \textit{trokin}, \textit{yercac1}, \textit{trsowl}, \textit{banana}}, and \textbf{\textit{wbwwre1}}.
 We selected species based on observation count and visualized to assess clustering behavior in the embedding space.
 }
 \label{fig:pacmap-perch}
\end{figure}

We build a baseline for the competition by utilizing pre-trained backbones for bioacoustic fauna classification.
The process of building a new classification head transfers the knowledge of the backbone to a new task, effectively reutilizing the representation space learned from the original task.
A domain-specific classifier for birds like Perch will be able to tightly cluster sounds that belong to specific classes like those in Figure~\ref{fig:pacmap-perch}.
These properties make pre-trained backbones desirable for similar domains.
We utilize the Bioacoustic Model Zoo to source our backbones, which serve as a companion to the OpenSoundscape project \cite{lapp2023opensoundscape}.
We enumerate the available models in Table~\ref{tab:model_zoo}.

\begin{table}[h]
\centering
\caption{
 Available Models in the Bioacoustics Model Zoo.
 We gather information from inference on a single 60-second test soundscape.
}
\label{tab:model_zoo}
\begin{tabular}{l|ccccc}
\hline
\textbf{Model Name} & \textbf{Clip Length (s)} & \textbf{Step Size (s)} & \textbf{Rows} & \textbf{Predict} & \textbf{Embed} \\ \hline
BirdNET & 3 & 3 & 20 & 6522 & 1024 \\
Perch & 5 & 5 & 12 & 10932 & 1280 \\
BirdSetConvNeXT & 5 & 5 & 12 & 9736 & 1024 \\
BirdSetEfficientNetB1 & 5 & 5 & 12 & 9736 & 1280 \\
RanaSierraeCNN & 2 & 2 & 30 & 2 & 512 \\
HawkEars & 3 & 3 & 20 & 333 & 2048 \\
YAMNet & 0.96 & 0.48 & 124 & 521 & 1024 \\
\hline
\end{tabular}
\end{table}

A large variety of models are available from the Bioacoustics Model Zoo, but they cannot be run directly in the competition for various reasons.
First, additional parameters and nodes in the computation graph needed for training can be pruned, fused, and optimized for inference.
We achieve an order of magnitude speedup for Perch when using TFLite over TensorFlow, enabling the model to run within competition time limits.

The other source of contention is the impedance mismatch between the clip length of models and the clip length expected by the competition.
The BirdNET model exemplifies this issue since it is lightweight enough to deploy as-is but requires adaptation for the competition.
In cases like this, we apply a sliding window over the 5-second target frame and extract embeddings from each window.
These embeddings can be aggregated by directly averaging them and applying a classification head or by running the classifier on each window individually and averaging the resulting probabilities.

To simplify model training, we first train the classifier on the original window embeddings.
During inference, we average the window embeddings over the target frame and run the classifier on the average embedding.
A window is within the target frame if it overlaps with it by at least 50\% of the window size, ensuring that each window contributes to only one target frame.
We plan to explore other aggregation methods in the future, such as max pooling; however, averaging works sufficiently well as a baseline.

\subsection{Spectrogram Token Skip-Gram Embeddings for Classification}

\begin{figure}
 \centering
 \includegraphics[width=1\linewidth]{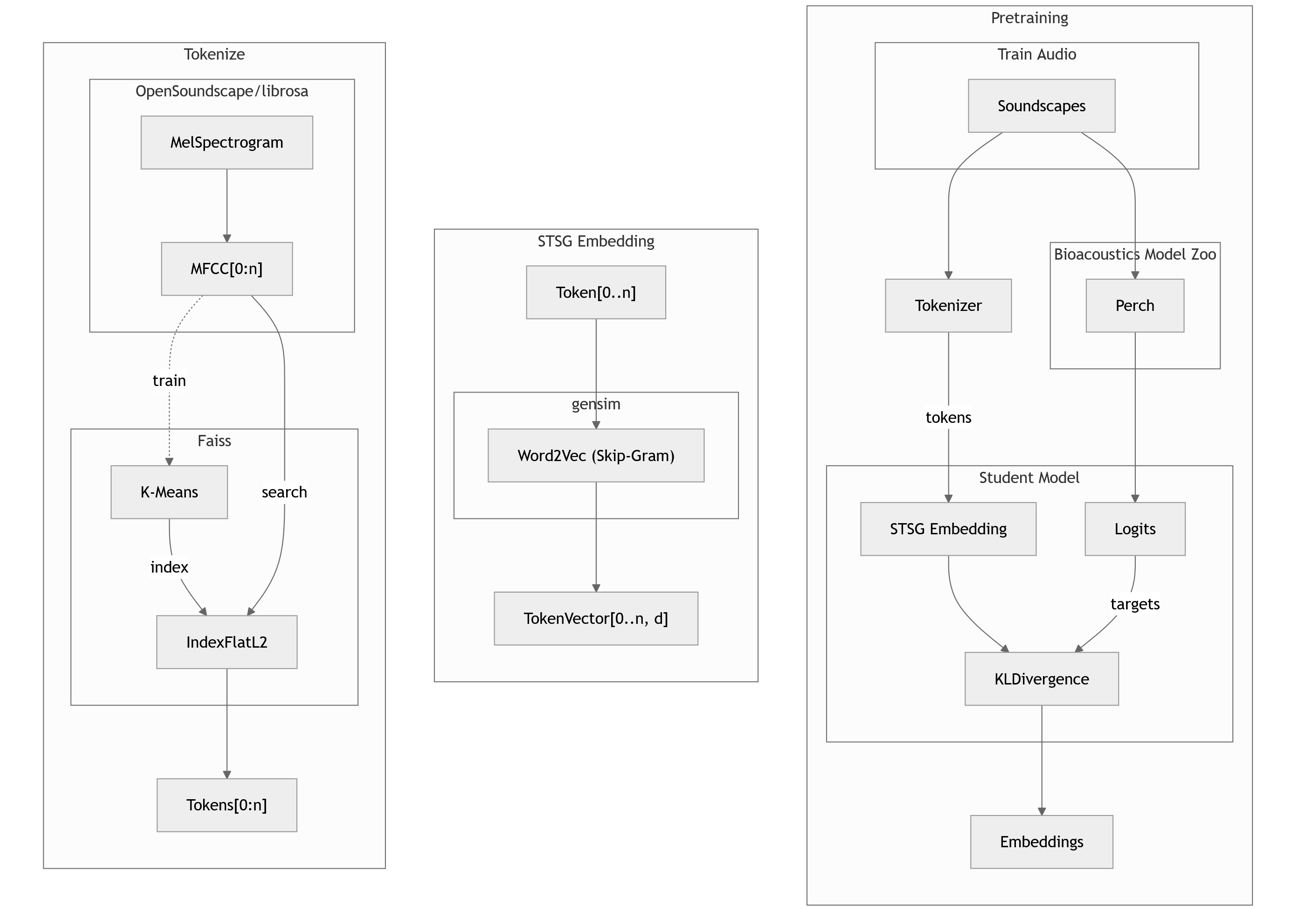}
 \caption{
 The Spectrogram Token Skip-Gram (STSG) Pipeline.
 Mel-spectrograms are generated from the source audio, which are then clustered using K-Means to create discrete tokens.
 A Word2Vec skip-gram model then learns a continuous vector embedding for each token in an unsupervised manner.
 In an optional pre-training stage, a student model uses these embeddings to produce output logits.
 A larger teacher model, Perch, also produces target logits from the same audio.
 The student's embeddings minimize the KL Divergence between its logits and the teacher's logits.
 }
 \label{fig:stsg-pipeline}
\end{figure}

The requirements for the Spectrogram Token Skip-Gram (STSG) embeddings are computationally efficient transformations that map raw waveforms into discrete token space and tokens into continuous embedding space.
One advantage of this representation is that we retain the sequential nature of audio, allowing us to contextualize a discrete token of audio with the surrounding tokens in time.
We approach this problem in three parts, as shown in Figure~\ref{fig:stsg-pipeline}, specifically through tokenization, embedding, and model pre-training.

\begin{table}[htbp]
\centering
\caption{
 Tuned hyperparameters for tokenization and sequence modeling.
 We also have a single parameter for the total number of tokens (codebook generation), which is the number of K-means clusters used to quantize the Mel-spectrogram.
}
\label{tab:pipeline_params_full}

\begin{subtable}{.35\textwidth}
\centering
\caption{Mel-spectrogram Feature Extraction}
\label{tab:mfcc_params_sub_full}
\begin{tabular}{lc}
\hline
\textbf{Parameter} & \textbf{Value} \\
\hline
Sample Rate & 32,000 Hz \\
Window Size & 8,000 samples \\
Overlap Fraction & 0.5 \\
Hop Size & 4,000 samples \\
Mel Bands & 768 \\
\hline
\textbf{Frames per Second} & \textbf{8} \\ \hline
\end{tabular}
\end{subtable}
\hfill 
\begin{subtable}{.6\textwidth}
\centering
\caption{Word2Vec Training}
\label{tab:word2vec_params_sub_full}
\begin{tabular}{lcl}
\hline
\textbf{Parameter} & \textbf{Value} & \textbf{Description} \\
\hline
\texttt{vector\_size} & 384 & Embedding dimensionality. \\
\texttt{sg} & 1 & Use Skip-gram algorithm. \\
\texttt{window} & 80 & Context size for co-occurrence. \\
\texttt{negative} & 5 & Negative samples for loss. \\
\texttt{ns\_exponent} & 0.0 & Negative sampling distribution. \\
\texttt{sample} & 1e-5 & Subsampling for frequent tokens. \\
\texttt{min\_count} & 1 & Pruning threshold for tokens. \\
\hline
\end{tabular}
\end{subtable}

\end{table}

\subsubsection{Spectrogram Tokenization via Faiss on MFCCs}

We choose a simple tokenizing scheme and sequence model to maximize inference throughput.
Our tokenizer clusters Mel-spectrograms across the training soundscape using K-means to generate a codebook for discrete representation \cite{davis1980comparison}.

We set the spectrogram parameters in Table~\ref{tab:mfcc_params_sub_full} in such a way that we obtain eight spectrogram frames per second; in other words, each resulting token will represent 0.125 seconds of context from the original spectrogram.
This value is chosen as a power of two for efficiency and is selected to yield a total sequence length of 40, corresponding to a 5-second prediction frame.
We can increase the tokens per second to increase our target sequence length but at the expense of losing temporal relevance at the spectrogram level.
We use a total of 768 Mel-frequency bands for testing.
This number is less than 10\% of the window size, which helps reduce the number of zero-valued frequency bands.

Once we choose parameters for the Mel-spectrograms, we compute them over the entire training soundscape dataset.
We perform principal component analysis (PCA) on the normalized Mel-spectrogram vectors to reduce their dimensionality for K-means and retrieval.
Normalizing the spectrogram vector helps reduce the effects of decibel intensities and instead focus on the distribution of the audio spectra.
PCA will project the spectrogram frames into a low-rank space, which helps clustering by reducing high-frequency noise and the overall computational and memory requirements.
We run K-means via Faiss over the frames to quantize them into integer tokens \cite{douze2024faiss}.

To obtain a new token, we generate the spectrogram frame for a given time interval, normalize it, and project it into a low-rank space. We then find the closest K-means vector that corresponds to it using the L2 distance.
The distance can be efficiently estimated through approximate nearest neighbor algorithms, such as those implemented in Faiss, including hierarchical navigable small worlds (HNSW).

\subsubsection{Skip-Gram Negative-Sampling Embeddings via Word2Vec} 

Once we have discrete tokens from the audio, we learn a model that embeds the token into a continuous space, which we can use for downstream classification tasks.
Our soundscape data lacks proper labels that we can utilize, but we can still learn about the relationship between tokens because audio is naturally sequential.
We achieve this in an unsupervised manner using a skip-gram negative-sampling embedding model, where we learn the location of a token in a high-dimensional space that captures temporal and semantic relevance within a sequential window.

We use the Word2Vec implementation in gensim to obtain a static lookup table for a learned skip-gram embedding model \cite{mikolov2013efficient} \cite{rehurek_lrec}.
We explore the initial parameter space in Table~\ref{tab:word2vec_params_sub_full}.
The objective is to learn a vector representation for a target token, $\vec{w}$, that is predictive of its actual context tokens, $\vec{c}$, while being dissimilar to $k$ negative samples, $\vec{c}_N$, drawn from the corpus. 
We minimize the loss function for each positive pair~\cite{caselles2018word2vec}:

$$
l_{\text{sgns}} = -\log(\sigma(\vec{w} \cdot \vec{c})) - \sum_{i=1}^{k} \log(\sigma(-\vec{w} \cdot \vec{c}_{N,i}))
$$

The negative samples are drawn from a unigram distribution based on token frequency $f(c)$ smoothed by hyperparameter $\alpha$ (the \texttt{ns\_exponent}) normalized over the corpus vocabulary:

$$
P(c) = \frac{f(c)^{\alpha}}{\sum_{c'} f(c')^{\alpha}}
$$

The value of $\alpha$ is a critical, task-dependent hyperparameter.
$\alpha=0.75$ is standard for NLP tasks.
A higher exponent exaggerates the sampling of frequent noise tokens, forcing the model to become highly discriminative between rare signal tokens and common background sounds. 
In contrast, a negative $\alpha$ would make the model focus on distinguishing rare signals from each other~\cite{caselles2018word2vec}.

We train a classification head on the resulting embeddings.
An embedding-centric workflow ensures that discrete token-based methods are comparable against vision methods over spectrograms.

\subsubsection{STSG Student-Teacher Pretraining}

In addition to learning a continuous representation of the spectrogram tokens, we experimented with student-teacher modeling to see if we could impart the bioacoustical domain knowledge from Perch into a smaller student model based on the STSG embedding space.
We configure the student with a 1D-CNN to aggregate STSG embeddings, project these into a latent space, and then attempt to match the probability distribution of Perch using KL-divergence as the loss.
We apply temperature scaling to the outputs such that teacher probabilities are softened by a squared factor of temperature $T = 3$.

\begin{table}[h!]
\centering
\caption{The architecture of our lightweight Student Model. The model takes a batch of token sequences of shape $(B, S)$ as input, where $B$ is the batch size and $S$ is the sequence length (40 tokens for a 5-second clip).}
\label{tab:student_model_arch}
\begin{tabular}{l l l}
\toprule
\textbf{Layer Type} & \textbf{Description} & \textbf{Output Shape} \\
\midrule
\multicolumn{2}{l}{\textit{\textbf{Input Token Sequence}}} & $(B, S)$ \\
Embedding & Vocabulary: 16,384, Dimension: 768 & $(B, S, 768)$ \\
\midrule
\multicolumn{3}{l}{\textit{\textbf{Aggregation Block}}} \\
Conv1d & In: 768, Out: 512, Kernel: 3, Pad: 1 & $(B, 512, S)$ \\
ReLU & Non-linear activation & $(B, 512, S)$ \\
BatchNorm1d & Stabilizes activations & $(B, 512, S)$ \\
AdaptiveMaxPool1d & Pools features across the sequence & $(B, 512, 1)$ \\
\midrule
\multicolumn{3}{l}{\textit{\textbf{Projection Head}}} \\
(Flatten) & Removes the last dimension & $(B, 512)$ \\
Linear & In: 512, Out: 512 & $(B, 512)$ \\
ReLU & Non-linear activation & $(B, 512)$ \\
BatchNorm1d & Stabilizes final activations & $(B, 512)$ \\
\midrule
\multicolumn{3}{l}{\textit{\textbf{Classification Head}}} \\
Linear & In: 512, Out: 10932 & $(B, 10932)$ \\
\midrule
\multicolumn{3}{l}{\textit{\textbf{Loss Function}}} \\
KLDivLoss & \multicolumn{2}{l}{Compares output to teacher model logits} \\
\bottomrule
\end{tabular}
\end{table}

We use the student model from Table~\ref{tab:student_model_arch} and train it against 80\% of the training soundscape, using the remaining 20\% as validation.
Then, we can pre-compute embeddings for the training species dataset, just as we do for other models in the bioacoustics model zoo.
In our modeling, we rely on complete sequences, i.e., masking is not implemented.

\subsection{Model Validation via Surrogate Classification Task}

To validate our transfer learning and token embedding models, we require a task that reflects the characteristics of the competition task.
We rely on the train species dataset in order to validate our models.
The surrogate task serves as a proxy for multi-label inference on soundscapes, for which we lack labeled data to work.
The training data skews in favor of commonly appearing species that are easier for individual people to crowd-source recordings.
We pre-compute vector representations of the audio aligned to the step size of the original classifier.
We drop species that have fewer than two samples, then stratify the dataset into test and validation in an 80/20 split.
We train a classification head on the embeddings with a hidden layer, non-linearity, and then cross-entropy on a linear projection to the output space.
We document the architecture of the classification head in Table~\ref{tab:linear_classifier_arch}.

\begin{table}[h!]
\centering
\caption{The architecture of the linear classifier head. This model takes a batch of pre-computed feature vectors of shape $(B, D_{in})$ as input, where $B$ is the batch size and $D_{in}$ is the input feature dimension. It outputs logits for $N_{classes}$ classes.}
\label{tab:linear_classifier_arch}
\begin{tabular}{l l l}
\toprule
\textbf{Layer Type} & \textbf{Description} & \textbf{Output Shape} \\
\midrule
\multicolumn{2}{l}{\textit{\textbf{Input Feature Vector}}} & $(B, D_{in})$ \\
\midrule
\multicolumn{3}{l}{\textit{\textbf{Classification Head}}} \\
Linear & In: $D_{in}$, Out: 512 & $(B, 512)$ \\
ReLU & Non-linear activation & $(B, 512)$ \\
Linear & In: 512, Out: $N_{classes}$ & $(B, N_{classes})$ \\
\midrule
\multicolumn{3}{l}{\textit{\textbf{Loss Function}}} \\
CrossEntropyLoss & \multicolumn{2}{l}{Standard loss for multi-class classification} \\
\bottomrule
\end{tabular}
\end{table}

We select a smaller semi-representative set of species to facilitate hyperparameter tuning sweeps over the STSG embeddings.
In Table~\ref{tab:validation_species}, we use Gemini 2.5 Pro (gemini-2.5-pro-preview-05-06 with Gemini App system prompting) to select species using provided taxonomy information, with explicit prompting about the region and competition to ensure mammals, insects, and amphibians have representation.
We validate the number of intervals processed by Perch and note the distribution across our small validation set.

\begin{table}[h!]
\centering
\caption{
 Species composition and embedding counts for the surrogate validation task. 
 The 20 species are diversified to represent different taxonomic classes.
}
\label{tab:validation_species}
\begin{tabular}{lllr}
\hline
\textbf{Primary Label} & \textbf{Common Name} & \textbf{Class} & \textbf{Count} \\
\hline
\texttt{socfly1} & Social Flycatcher & Aves & 3,347 \\
\texttt{strcuc1} & Striped Cuckoo & Aves & 2,862 \\
\texttt{tropar} & Tropical Parula & Aves & 2,578 \\
\texttt{strfly1} & Streaked Flycatcher & Aves & 2,443 \\
\texttt{butsal1} & Buff-throated Saltator & Aves & 2,282 \\
\texttt{52884} & True Crickets & Insecta & 1,191 \\
\texttt{blhpar1} & Blue-headed Parrot & Aves & 1,006 \\
\texttt{cattyr} & Cattle Tyrant & Aves & 808 \\
\texttt{eardov1} & Eared Dove & Aves & 500 \\
\texttt{neocor} & Neotropic Cormorant & Aves & 491 \\
\texttt{blkvul} & Black Vulture & Aves & 402 \\
\texttt{555086} & Rusty Tree Frog & Amphibia & 259 \\
\texttt{67252} & Veined Tree Frog & Amphibia & 231 \\
\texttt{savhaw1} & Savanna Hawk & Aves & 230 \\
\texttt{1462737} & Docidocercus fasciatus & Insecta & 140 \\
\texttt{65344} & Boettger's Colombian Tree Frog & Amphibia & 92 \\
\texttt{1192948} & Oxyprora surinamensis & Insecta & 83 \\
\texttt{41778} & Neotropical River Otter & Mammalia & 40 \\
\texttt{47067} & Brown-throated Three-toed Sloth & Mammalia & 14 \\
\texttt{42113} & Collared Peccary & Mammalia & 4 \\
\hline
\end{tabular}
\end{table}

The competition metric is a macro-averaged ROC-AUC that skips classes with no true positive labels.
We use the macro-averaged ROC-AUC as a rough proxy for our experiments.
This measure often fails to provide precise measurements, as it frequently remains near 1.0 in transfer learning experiments.
We will also report the F1-macro score, which enables more precise observations between models with different hyperparameters.
\section{Results}

\subsection{STSG Hyperparameter Tuning}

The STSG training process has a large number of parameters that can affect the output results.
For the sake of time, we select a set of hyperparameters that suffice for validating that the pipeline works in both training and inference and that it yields improvements over baseline methods.
Refer to Table~\ref{tab:mfcc_params_sub_full} for the set of default parameters.

\begin{table}[h!]
\centering
\caption{Key architectural and hyperparameter differences between the STSG model versions. The v1 model used MFCCs as its base feature, while the v2 series used tokens derived from a PCA-reduced Mel-spectrogram.}
\label{tab:model_versions}
\begin{tabular}{l l c c c c}
\toprule
\textbf{Model Version} & \textbf{Base Feature} & \textbf{Embedding} & \textbf{Context} & \textbf{NS Exponent} & \textbf{Subsampling} \\
\midrule
STSG (v1) & MFCC & 384 & 80 & 0.75 & \texttt{1e-4} \\
\midrule
STSG (v2.0) & PCA Mel-spectrogram & 1024 & 80 & 1.5 & \texttt{1e-5} \\
STSG (v2.1) & PCA Mel-spectrogram & 1024 & 80 & 0.0 & \texttt{1e-5} \\
STSG (v2.2) & PCA Mel-spectrogram & 1024 & 80 & -0.75 & \texttt{1e-4} \\
\bottomrule
\end{tabular}
\end{table}

Our first attempt at modeling used Mel-Frequency Cepstral Coefficients (MFCCs), which are effectively the discrete cosine transform (DCT) of the Mel-spectrogram.
We denote this STSG v1, with hyperparameter plots in Appendix~\ref{sec:stsg_v1}.
However, we artificially limit the representational power of the tokens in v1 by being overly conservative with the parameters of 128 Mel bands and 20 MFCCs.
In Table~\ref{tab:model_versions}, we describe the parameters for the v2 line of models that use the Mel-spectrogram representation of 768 Mel-bands.
The remaining hyperparameter results that follow utilize the Mel-spectrogram representation.

\begin{figure}[h!]
    \centering
    \includegraphics[width=0.7\columnwidth]{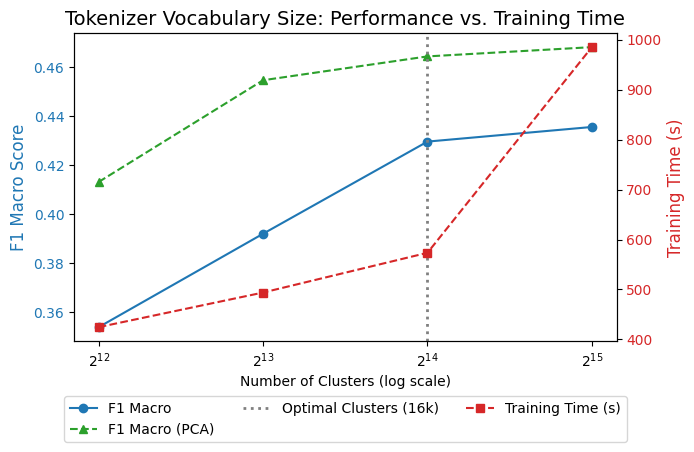}
    \caption{F1 score and training time vs. tokenizer vocabulary size. A vocabulary of 16k clusters provides the best performance before diminishing returns.}
    \label{fig:vocab_tradeoff_v2}
\end{figure}

Given a spectrogram representation, we must determine how many centroids we compute over the entire soundscape dataset.
The centroids act as a data-driven vocabulary for the audio corpus.
In Figure~\ref{fig:vocab_tradeoff_v2}, we sweep over the parameters of the vocabulary.
We settle on a token set size of 16k, which both fits inside a 2-byte integer and provides a reasonable amount of support, given that there are 3.73 million tokens in the 80\% training split of the soundscape.
While increasing the number of tokens to 32k improves the validation F1-macro scores, it also increases the overall required training time for the embedding proportionally.

\begin{table}[h!]
\centering
\caption{Cumulative variance explained by the number of principal components retained for the Mel-spectrogram features with 768 Mel bands.}
\label{tab:pca_variance}
\begin{tabular}{cc}
\toprule
\textbf{Number of PCA Components} & \textbf{Cumulative Variance Explained} \\
\midrule
32 & 77.20\% \\
64 & 81.85\% \\
128 & 87.02\% \\
256 & 93.12\% \\
512 & 98.34\% \\
\bottomrule
\end{tabular}
\end{table}

Applying PCA to the tokenizer is also a net positive to the model.
We choose 128 dimensions, as shown in Table~\ref{tab:pca_variance}, which retains 87\% of the original variance of the 768 Mel-spectrogram bands.
Reducing the number of dimensions increases modeling performance, likely due to the omission of high-frequency noise and latent modeling of sparse substructures in the original data.
Additionally, many of the algorithms that we use (K-means and approximate K-NN) have algorithmic complexity that is linearly proportional to the number of dimensions.
Reducing the number of dimensions allows us to fit more rows into memory and execute fewer instructions overall when tokenizing the signal.

\begin{table*}[h!]
\centering
\caption{
STSG (v2) hyperparameter sweep results on the surrogate validation task. All results are compared against a baseline configuration using a PCA-based tokenizer. Deltas ($\Delta$) are calculated relative to this baseline. The best-performing result in each scan group is highlighted in bold.
}
\label{tab:stsg_v2_sweep_pca_baseline}
\begin{tabular}{llccrrc}
\toprule
\textbf{Scan Group} & \textbf{Parameter} & \textbf{Value} & \textbf{F1 Macro} & \textbf{$\Delta$ F1} & \textbf{Time (s)} & \textbf{$\Delta$ Time (s)} \\
\midrule
\multirow{4}{*}{Baseline} 
    & \texttt{tokenizer} & PCA & 0.483 & +0.000 & 474 & +0 \\
    & \texttt{vector\_size} & 384 & & & & \\
    & \texttt{window} & 80 & & & & \\
    & \texttt{ns\_exponent} & 1.5 & & & & \\
    & \texttt{sample} & \textit{1e-5} & & & & \\
\midrule
\multirow{1}{*}{Varying \texttt{tokenizer}}
    & \texttt{tokenizer} & Standard & 0.452 & -0.031 & 1062 & +588 \\
\midrule
\multirow{4}{*}{Varying \texttt{vector\_size}} 
    & \texttt{vector\_size} & 128 & 0.395 & -0.088 & 396 & -78 \\
    & \texttt{vector\_size} & 256 & 0.467 & -0.016 & 459 & -15 \\
    & \texttt{vector\_size} & 512 & 0.488 & +0.005 & 548 & +74 \\
    & \texttt{vector\_size} & 1024 & \textbf{0.501} & \textbf{+0.018} & \textbf{836} & \textbf{+362} \\
\midrule
\multirow{2}{*}{Varying \texttt{window}} 
    & \texttt{window} & 40 & 0.470 & -0.013 & 362 & -112 \\
    & \texttt{window} & 120 & 0.481 & -0.002 & 576 & +102 \\
\midrule
\multirow{2}{*}{Varying \texttt{ns\_exponent}} 
    & \texttt{ns\_exponent} & -0.5 & 0.471 & -0.012 & 495 & +21 \\
    & \texttt{ns\_exponent} & 0.0 & 0.476 & -0.007 & 492 & +18 \\
\midrule
\multirow{2}{*}{Varying \texttt{sample}}
    & \texttt{sample} & \textit{1e-4} & 0.483 & +0.000 & 841 & +367 \\
    & \texttt{sample} & \textit{1e-6} & 0.439 & -0.044 & 281 & -193 \\
\bottomrule
\end{tabular}
\end{table*}

Once we have determined our token size, we conduct hyperparameter sweeps over the Gensim Word2Vec model, as shown in Table~\ref{tab:stsg_v2_sweep_pca_baseline}.
We vary over vector size, window size, negative sampling exponent, and sample rate.
The vector size determines the adequate capacity of the model and sweeps over 128, 256, 384, 512, and 1028 dimensions.
The dimension size is the most important parameter, as it directly affects the model's ability to learn complex relationships between tokens.
Increasing the vector size by a factor of three increases the F1-macro score by 0.018 but also increases the training time by a factor of 1.7.
We find that lowering the window size from the default of 80 decreases the score by 0.013, whereas increasing the window size does not significantly improve the score.
However, the training time is significantly affected by the window size, with a smaller window size reducing the training time by 112 seconds.

We find that sweeping over the negative sampling exponent $\alpha$ does not yield a clear pattern; however, in this particular sweep, setting $\alpha$ to 0.0 yields the best performance.
This parameter controls the distribution of negative samples, and a setting of 0.0 indicates a uniform distribution over the vocabulary.
A positive exponent means negative samples are more likely to be drawn from the more frequent tokens, while a negative exponent means negative samples are more likely to be drawn from the less frequent tokens.
Finally, we sweep over the subsampling rate, which controls the frequency of token sampling during training.
We find that decreasing the sample rate to 1e-4 does not change the score but increases the training time by a factor of two.
Increasing the sample rate to 1e-6 results in a significant drop in performance, as only the most frequent tokens are sampled, and the model is unable to learn from the less frequent ones.
These results help motivate the final choice of hyperparameters, which are balanced to provide a good trade-off between performance and training time.

\begin{figure}[h!]
    \centering
    \begin{subfigure}[b]{0.48\columnwidth}
        \includegraphics[width=\textwidth]{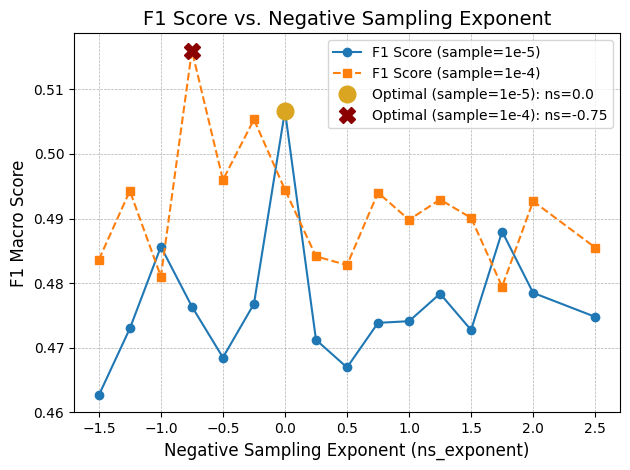}
        \caption{F1 score vs. ns\_exponent.}
        \label{fig:ns_exponent_tune_v2}
    \end{subfigure}
    \hfill 
    \begin{subfigure}[b]{0.48\columnwidth}
        \includegraphics[width=\textwidth]{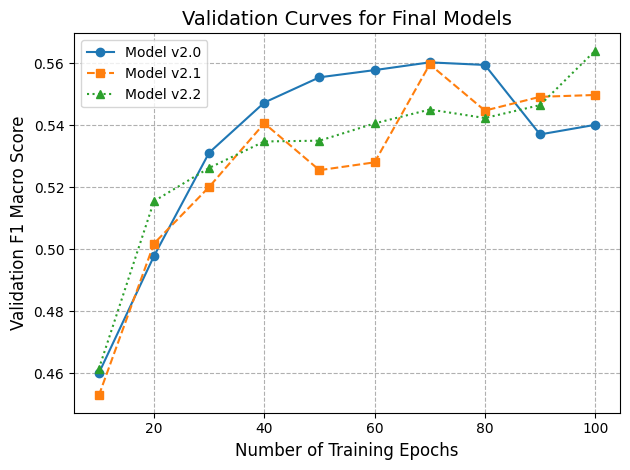}
        \caption{Word2Vec validation curves}
        \label{fig:comparison_curve_v2}
    \end{subfigure}
    \caption{
        Hyperparameter tuning and final model performance. (a) F1 score as a function of the negative sampling exponent (ns exponent) for two different subsampling rates. The optimal exponent shifts from 0.0 to -0.75 when using less aggressive subsampling (1e-4). 
        (b) Validation learning curves for the final STSG v2 models, showing that performance improves with more training but begins to overfit after approximately 70 epochs.
    }
    \label{fig:tuning_and_comparison_v2}
\end{figure}

Given the hyperparameter sweeps, we then fix over the parameters and run over the negative sampling exponent $\alpha$.
In our parameter scans, it was unclear how the negative sampling exponent and sample sizes interact with each other.
However, we find that when we set the sample to 1e-5, then, the best $\alpha$ is 0.0.
When we increase the number of samples by setting the sampling threshold to 1e-4, the best $\alpha$ decreases to -0.75.
When we aggressively subsample tokens in the former setting, we remove many of the most frequent tokens, which likely correspond to silence or background noise, making the resulting token distribution flatter.
The uniform strategy is effective since the noise is already downsampled.
When we are less aggressive with the samples, the distribution more closely resembles the original distribution.
Setting this to a negative value then forces the model to learn how to distinguish between tokens corresponding to different species rather than focusing on the frequent tokens.

We also look at the performance over time in Figure~\ref{fig:comparison_curve_v2} and find that 100 epochs are a reasonable length of time to run the Word2Vec algorithm.
In the first two iterations (v2.0 and v2.1), where we set subsampling to 1e-5, we observe either overfitting or significant stochasticity in the validation curve over time.
We increase the number of tokens available for training in v2.2, and this stabilizes training to a large degree.

\subsection{STSG Student Model Results}

\begin{table}[h!]
\centering
\caption{
    Overall model performance summary on the surrogate validation task, comparing the teacher model, a spectrogram baseline, and various versions of our token-based STSG and student models.
    }
\label{tab:overall_performance_summary}
\begin{tabular}{l ccc}
\toprule
\textbf{Model} & \textbf{Macro F1} & \textbf{Micro F1} & \textbf{Accuracy} \\
\midrule
MelSpec Baseline & 0.12 & 0.31 & 0.31 \\
Perch (Teacher) & 0.80 & 0.91 & 0.91 \\
\midrule
STSG (v1) & 0.45 & 0.53 & 0.53 \\
STSG (v2.0) & 0.54 & 0.60 & 0.60 \\
STSG (v2.1) & 0.55 & 0.60 & 0.60 \\
STSG (v2.2) & 0.56 & 0.61 & 0.61 \\
\midrule
STSG Student (v1) & 0.24 & 0.32 & 0.33 \\
STSG Student (v2.0) & 0.49 & 0.48 & 0.48 \\
STSG Student (v2.1) & 0.49 & 0.47 & 0.47 \\
STSG Student (v2.2) & 0.47 & 0.47 & 0.47 \\
\bottomrule
\end{tabular}
\end{table}

We report the results of the STSG student model in Table~\ref{tab:overall_performance_summary} for the validation task, comparing it to the Perch teacher model and our best token-based model.
These initial results show that the teacher model is competent, achieving a macro F1 score of 0.80 and a micro F1 score of 0.91.
The Mel-spectrogram, used directly by averaging the features over the time step, achieves a score of 0.12.
The STSG model performs better than the spectrogram baseline but worse than the teacher, with a score of 0.56. 
The token model can learn a representation space of the soundscape that can be transferred to the training dataset to a lower degree than the Perch model but is significantly better than choosing at random.

Finally, we observe that the student model, trained on the STSG representation space and then fine-tuned on the surrogate task, achieves a macro F1 score of 0.47.
The student model's performance represents a significant regression compared to using the STSG embeddings directly, suggesting that the STSG embeddings are unable to capture the complex relationships needed to represent the Perch logits effectively and that the geometry of the learned embedding space does not transfer well to classifying the training species dataset.

\subsection{Transfer Learning and Leaderboard Results}

\begin{table}[h]
\centering
\caption{
A final comparison of model performance on the surrogate classification task. We report F1 scores and the macro-averaged Area Under the Receiver Operating Characteristic Curve (ROC-AUC), which aligns with the competition's primary evaluation metric.
}
\label{tab:model_performance}
\begin{tabular}{l | c | ccc | ccc}
\toprule
\multicolumn{1}{c|}{\textbf{Model Name}} & \multicolumn{1}{c|}{\textbf{Epoch}} & \multicolumn{3}{c|}{\textbf{Validation Scores}} & \multicolumn{3}{c}{\textbf{Training Scores}} \\
\cmidrule(lr){3-5} \cmidrule(lr){6-8}
& & F1 Macro & F1 Micro & ROC-AUC & F1 Macro & F1 Micro & ROC-AUC \\
\midrule
BirdNET & 10 & 0.876 & 0.873 & 0.999 & 0.913 & 0.907 & 1.000 \\
BirdSetConvNeXT & 8 & 0.876 & 0.884 & 0.998 & 0.944 & 0.952 & 1.000 \\
Perch & 11 & 0.857 & 0.869 & 0.997 & 0.947 & 0.950 & 1.000 \\
BirdSetEfficientNetB1 & 7 & 0.776 & 0.819 & 0.996 & 0.831 & 0.849 & 0.998 \\
RanaSierraeCNN & 14 & 0.221 & 0.243 & 0.894 & 0.238 & 0.250 & 0.901 \\
\midrule
STSG (v1) & 19 & 0.381 & 0.394 & --- & 0.568 & 0.508 & --- \\
STSG (v2.0) & 14 & 0.563 & 0.511 & 0.975 & 0.580 & 0.544 & 0.979 \\
STSG (v2.1) & 14 & 0.560 & 0.515 & 0.975 & 0.581 & 0.546 & 0.978 \\
STSG (v2.2) & 14 & 0.573 & 0.526 & 0.976 & 0.592 & 0.559 & 0.978 \\
\bottomrule
\end{tabular}
\end{table}

\begin{table}[h]
\centering
\caption{
Final model performance (Public and Private Macro ROC-AUC Scores on Kaggle) and inference speed comparison. 
Projections are for a hypothetical test set of 700 soundscapes against the 90-minute (5400s) competition deadline.
}
\label{tab:final_comparison_final}
\begin{tabular}{l ccrr}
\toprule
\textbf{Model} & \textbf{Public} & \textbf{Private} & \textbf{Avg. (s) / File} & \textbf{Projected Time} \\
\midrule
BirdNET (1s step) & 0.719 & 0.718 & 6.2s & 4340s ($\sim$72 min) \\
BirdNET (2.5s step) & 0.715 & 0.721 & 2.48s & 1736s ($\sim$29 min) \\
Perch (TensorFlow) & - & - & 17.0s & 11900s ($\sim$3.3 hrs) \\
Perch (TFLite) & 0.729 & 0.711 & 1.4s & 980s ($\sim$16 min) \\
BirdSetEfficientNetB1 & \textbf{0.810} & \textbf{0.778} & 2.21s & 1547s ($\sim$26 min) \\
BirdSetConvNeXT & 0.768 & 0.756 & 5.11s & 3577s ($\sim$60 min) \\
RanaSierraeCNN & 0.617 & 0.643 & 4.86s & 3402s ($\sim$57 min) \\
\midrule
STSG (v1) & 0.478 & 0.474 & 0.4s & 280s ($\sim$5 min) \\
STSG (v2.0) & 0.548 & \textbf{0.538} & 0.5s & 350s ($\sim$6 min) \\
STSG (v2.1) & \textbf{0.559} & 0.520 & 0.5s & 350s ($\sim$6 min) \\
STSG (v2.2) & 0.534 & 0.513 & 0.5s & 350s ($\sim$6 min) \\
\bottomrule
\end{tabular}
\end{table}

We report the transfer learning surrogate task in Table~\ref{tab:model_performance}.
Here, we note that the ROC-AUC score does not help us disambiguate between the relative performance of models.
We note that the F1-macro score correlates well with the ranked performance of the models while providing larger differences in scores for individual observations.
The metric is also built into the classification report in the scikit-learn library.
We find that our best STSG models yield results around 0.56, whereas the best computer vision backbone models achieve results around 0.87.

We report the final model performance and timing on the leaderboard in Table~\ref{tab:final_comparison_final}.
We note that the primary metric that we are looking at here is the modified ROC-AUC scores, which are much lower than the ROC-AUC metrics in our surrogate task.
We note that the bioacoustic model zoo performs well for this simple transfer learning task without requiring complex modeling to address the domain shift in the datasets.
In comparison, the token embedding models achieve a modest score of around 0.56.

However, the STSG model is lightweight and fast.
It is three times as fast (0.5 seconds per file vs. 1.4 seconds per file) when compared to Perch on TFLite.
The Word2Vec modeling is weak but can be replaced by more robust and complex token embeddings without altering the runtime characteristics as long as the token embeddings are static.
Additionally, we found that many of the Torch-based bioacoustical models did not require any compilation or optimization to run within the competition bounds.
A significant amount of tuning can be done despite the computational constraints, as evident from the final leaderboard for the 2025 competition.

\section{Discussion}

\subsection{Transfer Learning}

\begin{figure}[!h]
    \centering
    \begin{subfigure}[b]{0.33\textwidth}
        \includegraphics[width=\linewidth]{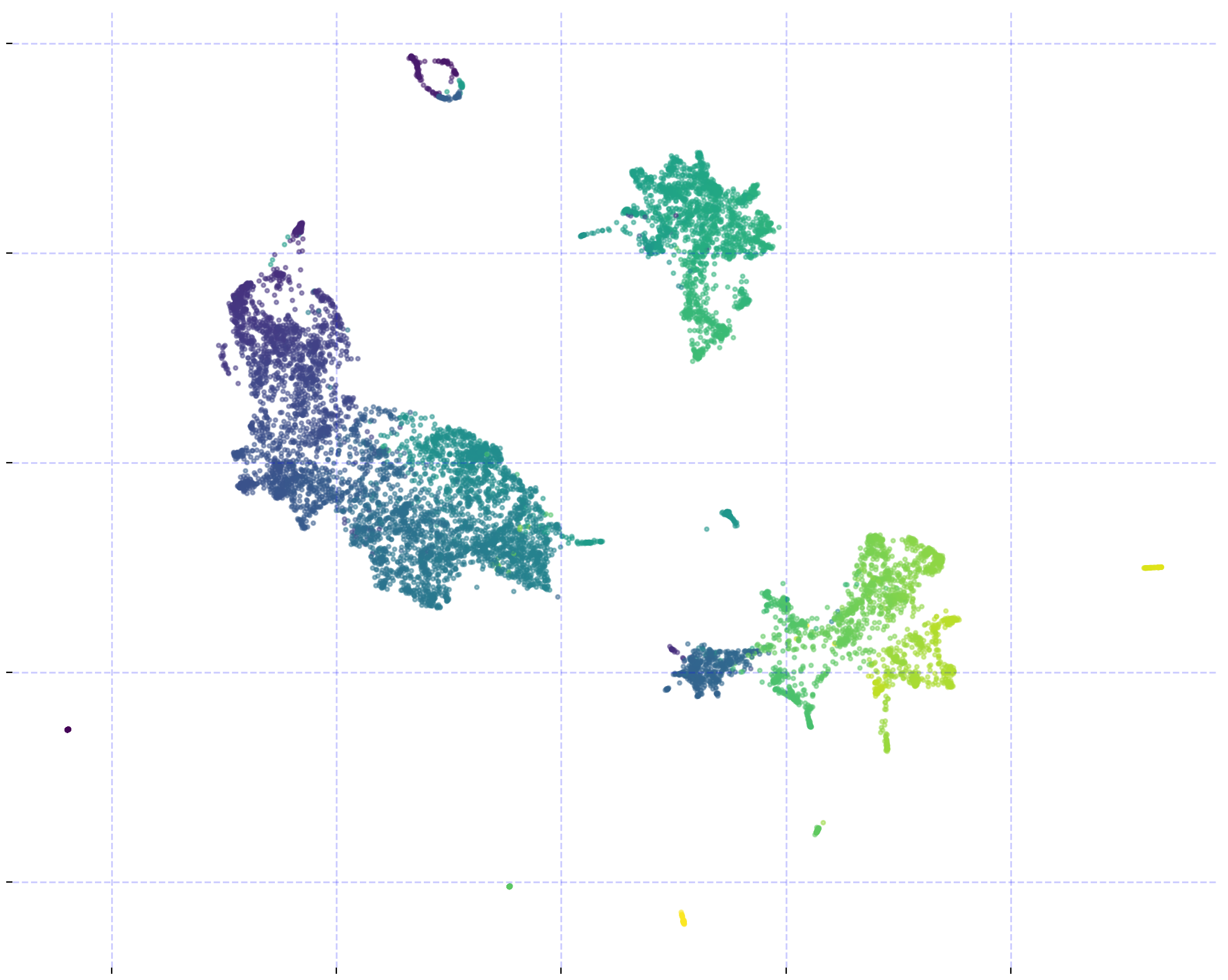}
        \caption{BirdNET.}
        \label{fig:image1}
    \end{subfigure}
    \hfill
    \begin{subfigure}[b]{0.33\textwidth}
        \includegraphics[width=\linewidth]{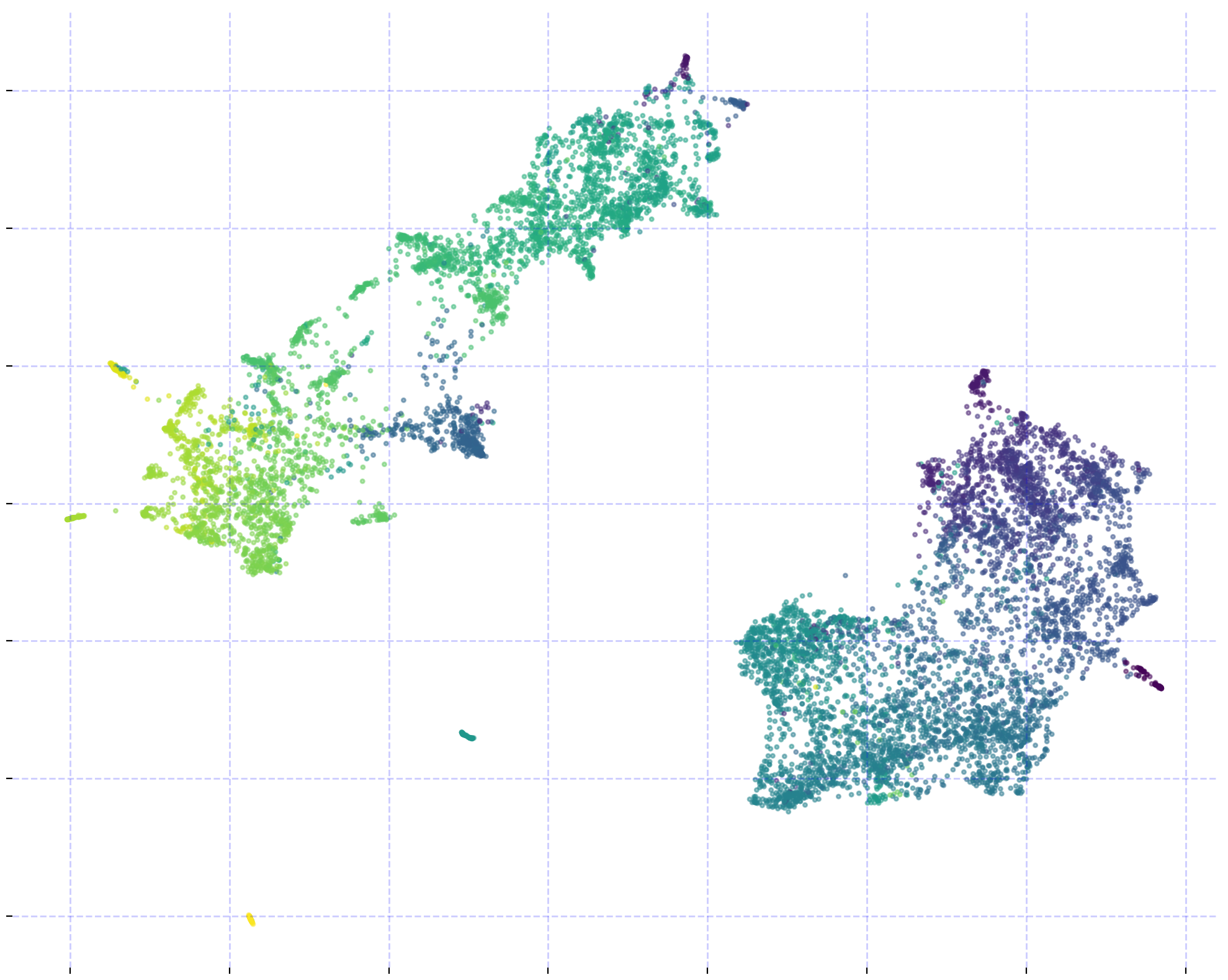}
        \caption{BirdSetConvNeXT.}
        \label{fig:image2}
    \end{subfigure}
    \hfill
    \begin{subfigure}[b]{0.33\textwidth}
        \includegraphics[width=\linewidth]{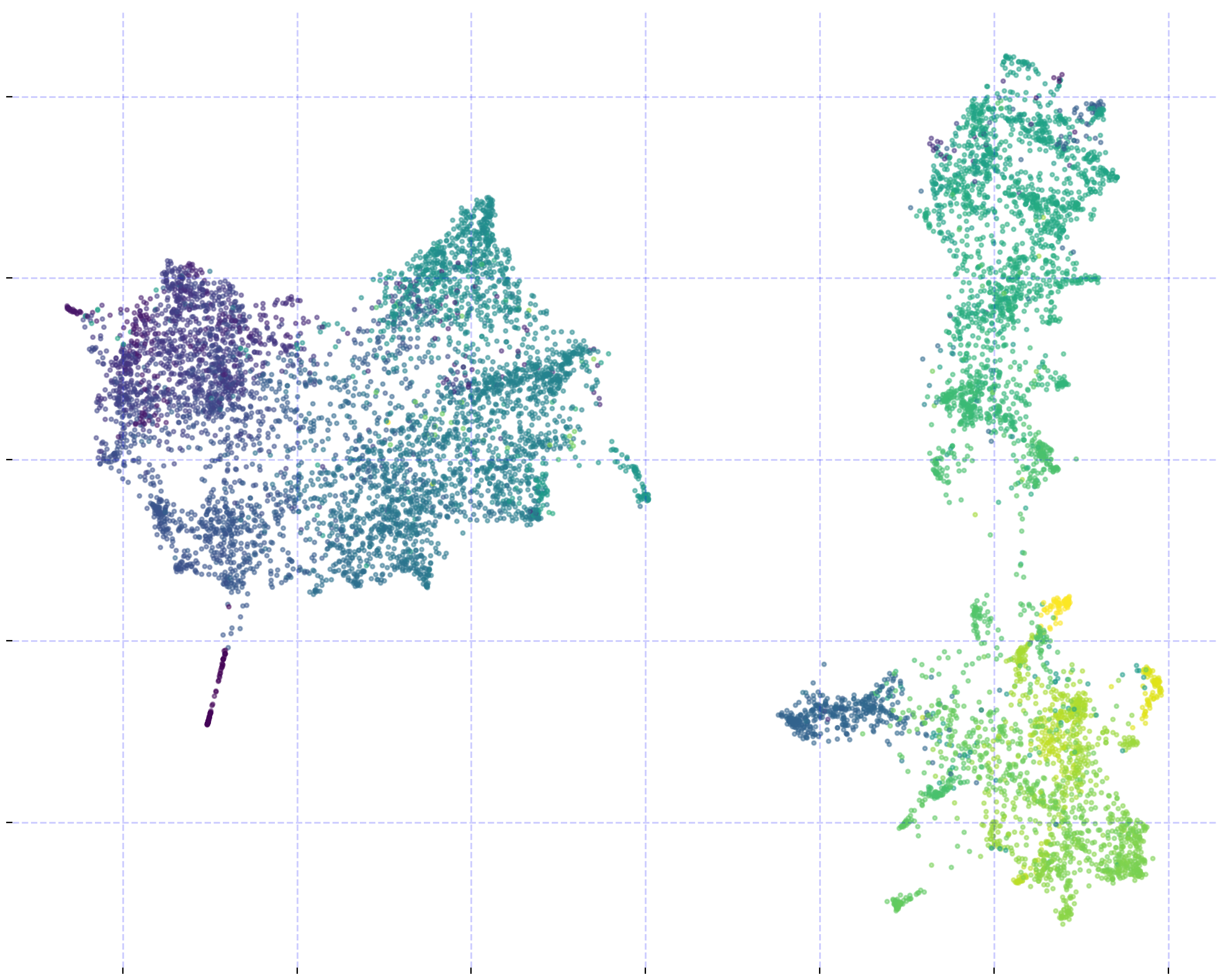}
        \caption{BirdSetEfficientNetB1.}
        \label{fig:image3}
    \end{subfigure}

    \vspace{0.5em}
    
    \begin{subfigure}[b]{0.33\textwidth}
        \includegraphics[width=\linewidth]{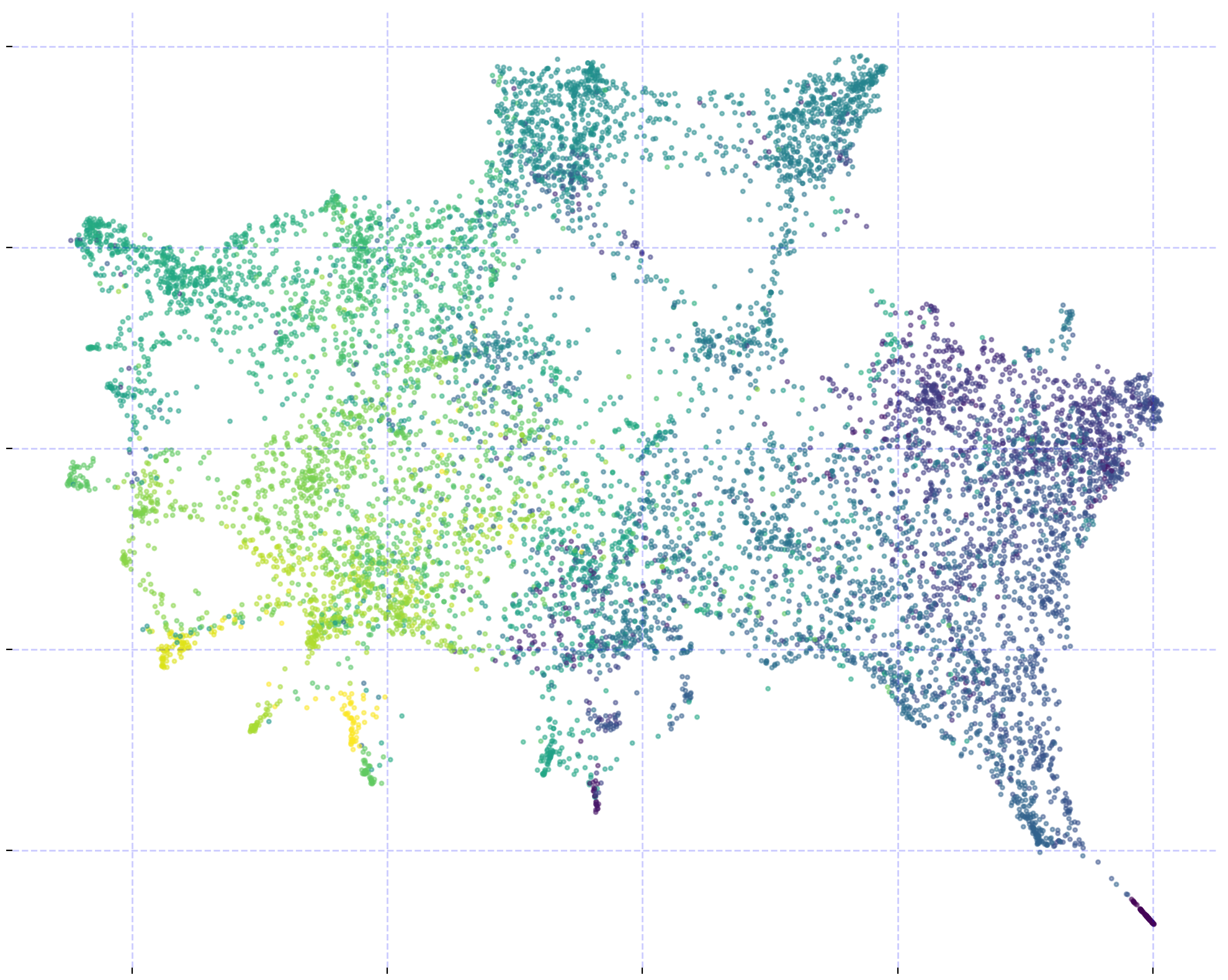}
        \caption{HawkEars.}
        \label{fig:image4}
    \end{subfigure}
    \hfill
    \begin{subfigure}[b]{0.33\textwidth}
        \includegraphics[width=\linewidth]{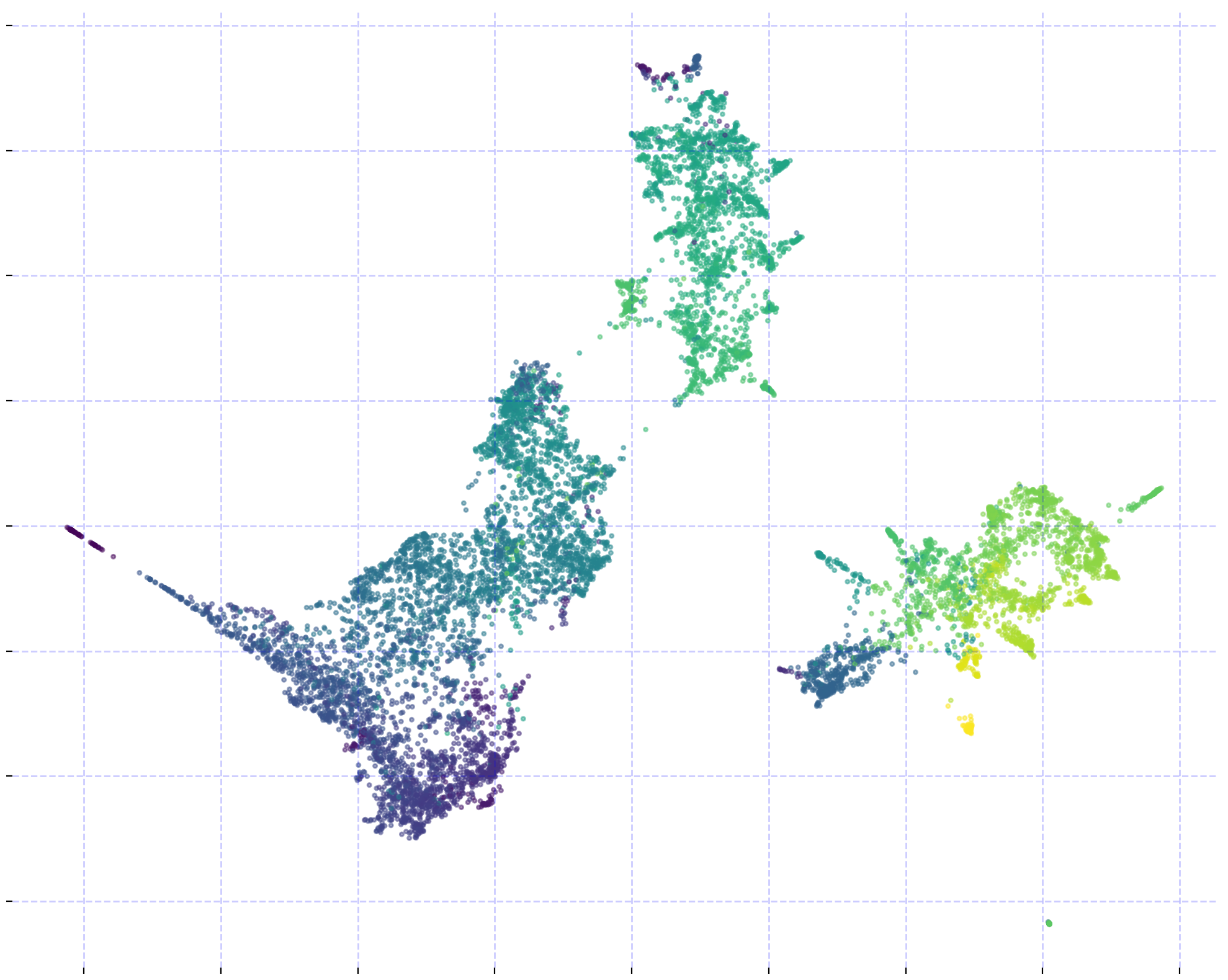}
        \caption{Perch.}
        \label{fig:image5}
    \end{subfigure}
    \hfill
    \begin{subfigure}[b]{0.33\textwidth}
        \includegraphics[width=\linewidth]{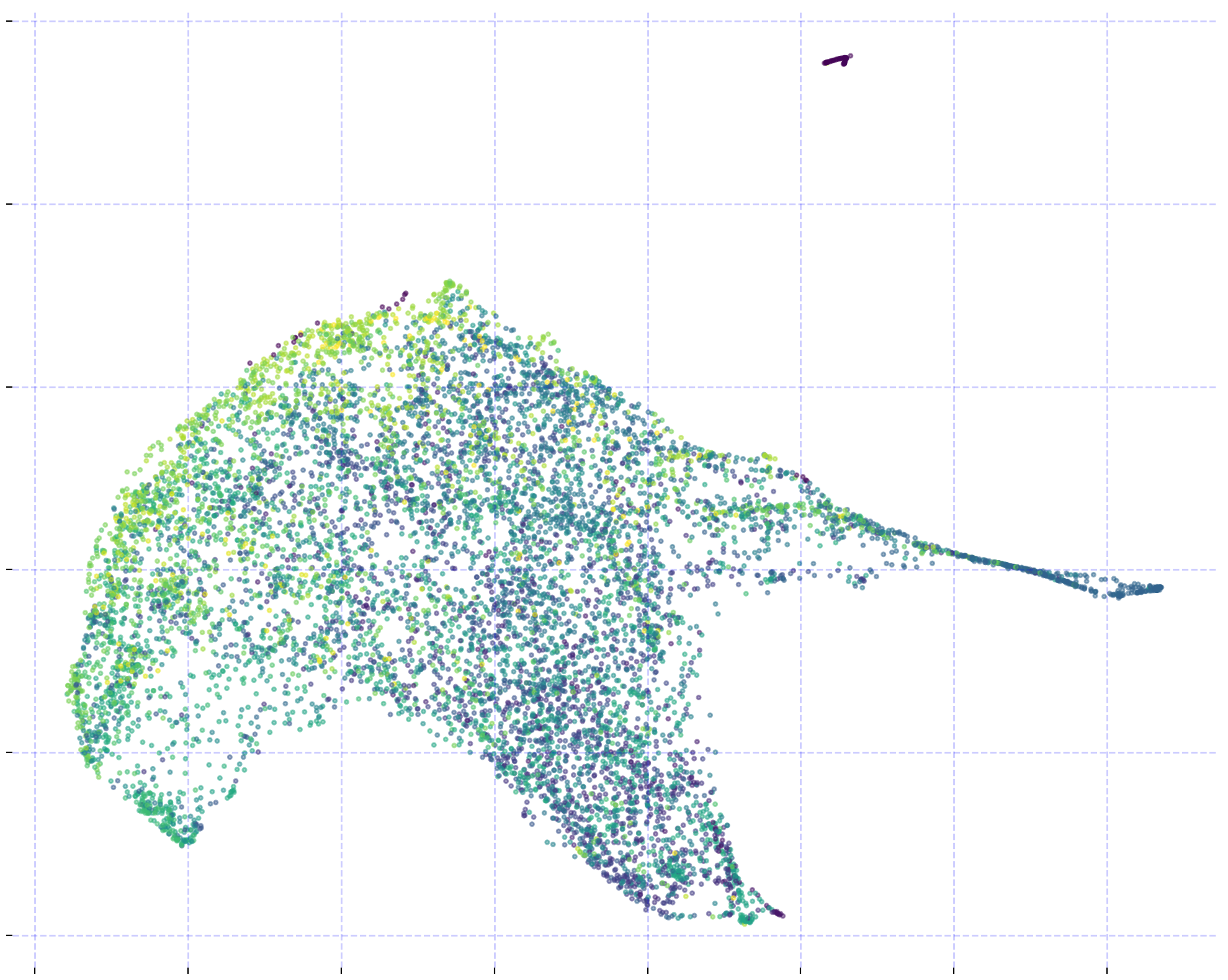}
        \caption{RanaSierraeCNN.}
        \label{fig:image6}
    \end{subfigure}

    \caption{
    PaCMAP visualizations of bioacoustic model zoo embedding spaces using the training dataset where every point represents an audio track.
    The color represents the one-dimensional embedding of the audio track in BirdNET space, computed by averaging all embeddings generated for the track.
    }
    \label{fig:2x3_grid}
\end{figure}

While the transfer learning experiments we have run are methodologically simple, we were able to test a much wider variety of models than in previous years.
The bioacoustic model zoo is an excellent library for development, as it provides a reasonable API into many popular bioacoustic backbones.
We visualize the embedding space extracted from the library in Figure~\ref{fig:2x3_grid}, which provides a qualitative view of the geometric structure of the backbones.
The tight clustering behavior of BirdNET, Perch, and BirdSet models is desirable, as it facilitates more straightforward mapping to our final objective space.
The color of the data corresponds to the one-dimensional projection of the embedding in BirdNET, which effectively ranks data points on a number line.
We expect to see a smooth transition in color based on distance in two dimensions.
The discoloration in RanaSierraCNN reflects the leaderboard performance and the decrease in performance due to ambiguity between similar points.
HawkEars is a model that we were unable to test on the leaderboard due to its ensembling structure; however, the domain-specific nature of Hawks leads to a more substantial domain shift, also observed in the Rana Sierra Frog classifier.
We also note that we did not test YAMNet due to the unusual clip length, although it may be helpful as a model for filtering.

BirdNET and Perch continue to be staple models that provide state-of-the-art performance with broad reach in real deployments.
The BirdSet dataset has made it much easier to develop new backbone models that rival the performance of BirdNET and Perch while staying in the Torch ecosystem and with prediction windows corresponding to the BirdCLEF challenge \cite{rauch2024birdset}.
The Tensorflow software stack complicates experiments where the primary modeling tool is Torch.
In particular, we encountered complications while running models in the bioacoustic model zoo that relied on both TensorFlow and Torch, which required identifying the appropriate version pins for inference.
Surprisingly, the BirdSetEfficientNetB1 model outperformed the TensorFlow models, achieving the best score on the public leaderboard at 0.81.
The BirdSet dataset and resulting models will likely form the basis of strong solutions in future iterations of the competition due to the ease of benchmarking and tooling.

\subsection{Spectrogram Tokens and Sequence Modeling}

In our experiment, we tested the ability to model bioacoustic data as discrete sequences and to determine if it could serve as a basis for student-teacher modeling.
We discovered a fast and lightweight token representation for our data that enables unsupervised modeling despite its limited discriminative power. 
Spectrogram tokens are an interesting result because they rely on algorithmic building blocks of clustering and nearest-neighbor search to represent the sequence sparsely.
The information bottleneck in our process lies in the signal representation given by the spectrogram.
When we increased the total number of Mel-bands and avoided dimensionality reduction early in the pipeline, our validation scores increased by approximately 0.1 in F1-macro.
A better representation leads to improved tokenization, enabling the sequence model to better represent the audio for classification purposes.

The idea of quantizing audio for downstream applications is not unique to this work.
There are data-driven neural codecs for compressing and decompressing audio, such as EnCodec \cite{defossez2022high}.
However, our team's initial explorations into neural codecs in the 2024 competition placed EnCodec specifically in the non-viable category due to the computational overhead of attention \cite{miyaguchi2024transfer}.
Other frameworks, such as wav2vec 2.0, closely mirror our pipeline but rely on a latent audio representation via CNNs that are quantized and then contextualized by Transformers.
Wav2vec would likely not work without modification for the BirdCLEF competition due to imposed inference deadlines.
The CNN for representing speech is likely viable due to the overwhelming presence of CNNs already. 
However, the Transformer network would need an efficient approximation at inference time.

\section{Future Work}

Future work will place greater emphasis on modeling complex sequences.
In particular, we would like to see better token representations and the use of Transformers to contextualize sequences of tokens.
It would be interesting to see how much of the wav2vec framework applies to BirdCLEF and to see whether contextualized tokens via Transformers can run on constrained systems.

While the STSG representation is fast, it leaves significant room for improvement.
We can use up most of the inference budget to calculate high-quality tokens, assuming we apply a static lookup table with a lightweight classification head.
The first direction is to replace the spectrogram token with a YAMNet-based token, which, although resulting in a shorter 2-token-per-second sequence, provides a more powerful representation that will form more meaningful clusters under K-means.
Another approach might be to use a clustering algorithm, such as HDBSCAN, to identify centroids that are better suited to the topology.
However, we lose the performance characteristics from having a highly optimized ANN algorithm in $L^p$-space.
The final direction is to learn tokens using wav2vec and to optimize the framework for CPU inference heavily.
Wav2vec is promising, although it requires a significant amount of engineering work to execute.

One of the factors that make transformer-based models challenging to use in the BirdCLEF+ competition is that inference is expensive with naive implementations due to the sheer number of parameters involved.
However, transformers are a natural way to represent sequential data because attention efficiently contextualizes tokens within sequences \cite{vaswani2017attention}.
We may be able to statically capture token representations learned by parameterization of transformer networks that are far larger than can be run on an inference machine by using compilation methods such as those found in Model2Vec \cite{minishlab2024model2vec}.
This technique establishes a static mapping of tokens by embedding single-token sequences, compressing them using PCA, and applying an information-based reweighting scheme that assumes a Zipf distribution of tokens.
These tokens can then be averaged across sentences and have demonstrated good performance on encoder-only, downstream tasks.

\section{Conclusions}

Through this work, we address computational limitations in bioacoustical classification by leveraging transfer learning of domain-specific classifiers and a lightweight token-based approach.
We demonstrate state-of-the-art models and their performance on the domain-shifted data through models like BirdNET and Perch in the bioacoustic model zoo.
Our best model was BirdSetEfficientNetB1, with a private ROC-AUC score of 0.778 and a public score of 0.810.
We also investigate the foundations of distilling complex sequence models into a static lookup of contextualized tokens.
Our Spectrogram Token Skip-Gram (STSG) pipeline achieves a runtime of 6 minutes on the test dataset, yielding a final ROC-AUC private score of 0.520 and a public score of 0.559.
The STSG pipeline provides the basis for more complex sequence models, as it enables the discovery of a representational space that can be parsed and transformed efficiently on a CPU.
Although the overall performance is low, there are interesting directions that this research can take, given that spectrogram tokens possess representational power.

Supporting code for this paper is located at \url{https://github.com/dsgt-arc/birdclef-2025}.

\section*{Acknowledgements}

We thank the Data Science at Georgia Tech (DS@GT) CLEF competition group for their support.
This research was supported in part through research cyberinfrastructure resources and services provided by the Partnership for an Advanced Computing Environment (PACE) at the Georgia Institute of Technology, Atlanta, Georgia, USA \cite{PACE}. 

\section*{Declaration on Generative AI}

During the preparation of this work, the author(s) used Gemini Pro in order to: Abstract drafting, formatting assistance, grammar and spelling check. After using these tool(s)/service(s), the author(s) reviewed and edited the content as needed and take(s) full responsibility for the publication’s content. 
\bibliography{main}

\begin{thebibliography}{20}
\expandafter\ifx\csname natexlab\endcsname\relax\def\natexlab#1{#1}\fi
\providecommand{\url}[1]{\texttt{#1}}
\providecommand{\href}[2]{#2}
\providecommand{\path}[1]{#1}
\providecommand{\DOIprefix}{doi:}
\providecommand{\ArXivprefix}{arXiv:}
\providecommand{\URLprefix}{URL: }
\providecommand{\Pubmedprefix}{pmid:}
\providecommand{\doi}[1]{\href{http://dx.doi.org/#1}{\path{#1}}}
\providecommand{\Pubmed}[1]{\href{pmid:#1}{\path{#1}}}
\providecommand{\bibinfo}[2]{#2}
\ifx\xfnm\relax \def\xfnm[#1]{\unskip,\space#1}\fi
\bibitem[{Cañas et~al.(2025)Cañas, Kahl, Denton, Toro-Gómez, Rodriguez-Buritica, Benavides-Lopez, Ulloa, Caycedo-Rosales, Klinck, Glotin, Go{\"e}au, Vellinga, Planqu{\'e}, and Joly}]{birdclef2025}
\bibinfo{author}{J.~S. Cañas}, \bibinfo{author}{S.~Kahl}, \bibinfo{author}{T.~Denton}, \bibinfo{author}{M.~P. Toro-Gómez}, \bibinfo{author}{S.~Rodriguez-Buritica}, \bibinfo{author}{J.~L. Benavides-Lopez}, \bibinfo{author}{J.~S. Ulloa}, \bibinfo{author}{P.~Caycedo-Rosales}, \bibinfo{author}{H.~Klinck}, \bibinfo{author}{H.~Glotin}, \bibinfo{author}{H.~Go{\"e}au}, \bibinfo{author}{W.-P. Vellinga}, \bibinfo{author}{R.~Planqu{\'e}}, \bibinfo{author}{A.~Joly},
\newblock \bibinfo{title}{Overview of {BirdCLEF+} 2025: Multi-taxonomic sound identification in the middle magdalena valley, colombia},
\newblock in: \bibinfo{booktitle}{Working Notes of CLEF 2025 - Conference and Labs of the Evaluation Forum}, \bibinfo{year}{2025}.
\bibitem[{Picek et~al.(2025)Picek, Kahl, Go{\"e}au, Adam et~al.}]{lifeclef2025}
\bibinfo{author}{L.~Picek}, \bibinfo{author}{S.~Kahl}, \bibinfo{author}{H.~Go{\"e}au}, \bibinfo{author}{L.~Adam}, et~al.,
\newblock \bibinfo{title}{Overview of lifeclef 2025: Challenges on species presence prediction and identification, and individual animal identification},
\newblock in: \bibinfo{booktitle}{International Conference of the Cross-Language Evaluation Forum for European Languages}, \bibinfo{organization}{Springer}, \bibinfo{year}{2025}.
\bibitem[{Ghani et~al.(2023)Ghani, Denton, Kahl, and Klinck}]{ghani2023global}
\bibinfo{author}{B.~Ghani}, \bibinfo{author}{T.~Denton}, \bibinfo{author}{S.~Kahl}, \bibinfo{author}{H.~Klinck},
\newblock \bibinfo{title}{Global birdsong embeddings enable superior transfer learning for bioacoustic classification},
\newblock \bibinfo{journal}{Scientific Reports} \bibinfo{volume}{13} (\bibinfo{year}{2023}) \bibinfo{pages}{22876}.
\bibitem[{Williams et~al.(2025)Williams, van Merri{\"e}nboer, Dumoulin, Hamer, Fleishman, McKown, Munger, Rice, Lillis, White et~al.}]{williams2025using}
\bibinfo{author}{B.~Williams}, \bibinfo{author}{B.~van Merri{\"e}nboer}, \bibinfo{author}{V.~Dumoulin}, \bibinfo{author}{J.~Hamer}, \bibinfo{author}{A.~B. Fleishman}, \bibinfo{author}{M.~McKown}, \bibinfo{author}{J.~Munger}, \bibinfo{author}{A.~N. Rice}, \bibinfo{author}{A.~Lillis}, \bibinfo{author}{C.~White}, et~al.,
\newblock \bibinfo{title}{Using tropical reef, bird and unrelated sounds for superior transfer learning in marine bioacoustics},
\newblock \bibinfo{journal}{Philosophical Transactions B} \bibinfo{volume}{380} (\bibinfo{year}{2025}) \bibinfo{pages}{20240280}.
\bibitem[{Kahl et~al.(2024)Kahl, Denton, Klinck, Ramesh, Joshi, Srivathsa, Anand, Arvind, CP, Sawant, Robin, Glotin, Go{\"e}au, Vellinga, Planqu{\'e}, and Joly}]{birdclef2024}
\bibinfo{author}{S.~Kahl}, \bibinfo{author}{T.~Denton}, \bibinfo{author}{H.~Klinck}, \bibinfo{author}{V.~Ramesh}, \bibinfo{author}{V.~Joshi}, \bibinfo{author}{M.~Srivathsa}, \bibinfo{author}{A.~Anand}, \bibinfo{author}{C.~Arvind}, \bibinfo{author}{H.~CP}, \bibinfo{author}{S.~Sawant}, \bibinfo{author}{V.~V. Robin}, \bibinfo{author}{H.~Glotin}, \bibinfo{author}{H.~Go{\"e}au}, \bibinfo{author}{W.-P. Vellinga}, \bibinfo{author}{R.~Planqu{\'e}}, \bibinfo{author}{A.~Joly},
\newblock \bibinfo{title}{Overview of {BirdCLEF} 2024: Acoustic identification of under-studied bird species in the western ghats},
\newblock \bibinfo{journal}{Working Notes of CLEF 2024 - Conference and Labs of the Evaluation Forum}  (\bibinfo{year}{2024}).
\bibitem[{Miyaguchi et~al.(2024)Miyaguchi, Cheung, Gustineli, and Kim}]{miyaguchi2024transfer}
\bibinfo{author}{A.~Miyaguchi}, \bibinfo{author}{A.~Cheung}, \bibinfo{author}{M.~Gustineli}, \bibinfo{author}{A.~Kim},
\newblock \bibinfo{title}{Transfer learning with pseudo multi-label birdcall classification for ds@gt birdclef 2024},
\newblock \bibinfo{journal}{Working Notes of CLEF 2024 - Conference and Labs of the Evaluation Forum}  (\bibinfo{year}{2024}).
\bibitem[{Zeghidour et~al.(2021)Zeghidour, Luebs, Omran, Skoglund, and Tagliasacchi}]{zeghidour2021soundstream}
\bibinfo{author}{N.~Zeghidour}, \bibinfo{author}{A.~Luebs}, \bibinfo{author}{A.~Omran}, \bibinfo{author}{J.~Skoglund}, \bibinfo{author}{M.~Tagliasacchi},
\newblock \bibinfo{title}{Soundstream: An end-to-end neural audio codec},
\newblock \bibinfo{journal}{CoRR} \bibinfo{volume}{abs/2107.03312} (\bibinfo{year}{2021}). \URLprefix \url{https://arxiv.org/abs/2107.03312}. \href{http://arxiv.org/abs/2107.03312}{{\tt arXiv:2107.03312}}.
\bibitem[{D{\'e}fossez et~al.(2022)D{\'e}fossez, Copet, Synnaeve, and Adi}]{defossez2022high}
\bibinfo{author}{A.~D{\'e}fossez}, \bibinfo{author}{J.~Copet}, \bibinfo{author}{G.~Synnaeve}, \bibinfo{author}{Y.~Adi},
\newblock \bibinfo{title}{High fidelity neural audio compression},
\newblock \bibinfo{journal}{arXiv preprint arXiv:2210.13438}  (\bibinfo{year}{2022}).
\bibitem[{Baevski et~al.(2020)Baevski, Zhou, Mohamed, and Auli}]{baevski2020wav2vec}
\bibinfo{author}{A.~Baevski}, \bibinfo{author}{Y.~Zhou}, \bibinfo{author}{A.~Mohamed}, \bibinfo{author}{M.~Auli},
\newblock \bibinfo{title}{wav2vec 2.0: A framework for self-supervised learning of speech representations},
\newblock \bibinfo{journal}{Advances in neural information processing systems} \bibinfo{volume}{33} (\bibinfo{year}{2020}) \bibinfo{pages}{12449--12460}.
\bibitem[{Mikolov et~al.(2013)Mikolov, Chen, Corrado, and Dean}]{mikolov2013efficient}
\bibinfo{author}{T.~Mikolov}, \bibinfo{author}{K.~Chen}, \bibinfo{author}{G.~Corrado}, \bibinfo{author}{J.~Dean},
\newblock \bibinfo{title}{Efficient estimation of word representations in vector space},
\newblock \bibinfo{journal}{arXiv preprint arXiv:1301.3781}  (\bibinfo{year}{2013}).
\bibitem[{Borsos et~al.(2023)Borsos, Marinier, Vincent, Kharitonov, Pietquin, Sharifi, Roblek, Teboul, Grangier, Tagliasacchi, and Zeghidour}]{borsos2023audiolm}
\bibinfo{author}{Z.~Borsos}, \bibinfo{author}{R.~Marinier}, \bibinfo{author}{D.~Vincent}, \bibinfo{author}{E.~Kharitonov}, \bibinfo{author}{O.~Pietquin}, \bibinfo{author}{M.~Sharifi}, \bibinfo{author}{D.~Roblek}, \bibinfo{author}{O.~Teboul}, \bibinfo{author}{D.~Grangier}, \bibinfo{author}{M.~Tagliasacchi}, \bibinfo{author}{N.~Zeghidour},
\newblock \bibinfo{title}{Audiolm: A language modeling approach to audio generation},
\newblock \bibinfo{journal}{IEEE/ACM Trans. Audio, Speech and Lang. Proc.} \bibinfo{volume}{31} (\bibinfo{year}{2023}) \bibinfo{pages}{2523–2533}. \URLprefix \url{https://doi.org/10.1109/TASLP.2023.3288409}. \DOIprefix\doi{10.1109/TASLP.2023.3288409}.
\bibitem[{Lapp et~al.(2023)Lapp, Rhinehart, Freeland-Haynes, Khilnani, Syunkova, and Kitzes}]{lapp2023opensoundscape}
\bibinfo{author}{S.~Lapp}, \bibinfo{author}{T.~Rhinehart}, \bibinfo{author}{L.~Freeland-Haynes}, \bibinfo{author}{J.~Khilnani}, \bibinfo{author}{A.~Syunkova}, \bibinfo{author}{J.~Kitzes},
\newblock \bibinfo{title}{Opensoundscape: an open-source bioacoustics analysis package for python},
\newblock \bibinfo{journal}{Methods in Ecology and Evolution} \bibinfo{volume}{14} (\bibinfo{year}{2023}) \bibinfo{pages}{2321--2328}.
\bibitem[{Davis and Mermelstein(1980)}]{davis1980comparison}
\bibinfo{author}{S.~Davis}, \bibinfo{author}{P.~Mermelstein},
\newblock \bibinfo{title}{Comparison of parametric representations for monosyllabic word recognition in continuously spoken sentences},
\newblock \bibinfo{journal}{IEEE Transactions on Acoustics, Speech, and Signal Processing} \bibinfo{volume}{28} (\bibinfo{year}{1980}) \bibinfo{pages}{357--366}. \DOIprefix\doi{10.1109/TASSP.1980.1163420}.
\bibitem[{Douze et~al.(2024)Douze, Guzhva, Deng, Johnson, Szilvasy, Mazaré, Lomeli, Hosseini, and Jégou}]{douze2024faiss}
\bibinfo{author}{M.~Douze}, \bibinfo{author}{A.~Guzhva}, \bibinfo{author}{C.~Deng}, \bibinfo{author}{J.~Johnson}, \bibinfo{author}{G.~Szilvasy}, \bibinfo{author}{P.-E. Mazaré}, \bibinfo{author}{M.~Lomeli}, \bibinfo{author}{L.~Hosseini}, \bibinfo{author}{H.~Jégou},
\newblock \bibinfo{title}{The faiss library}  (\bibinfo{year}{2024}). \href{http://arxiv.org/abs/2401.08281}{{\tt arXiv:2401.08281}}.
\bibitem[{{\v R}eh{\r u}{\v r}ek and Sojka(2010)}]{rehurek_lrec}
\bibinfo{author}{R.~{\v R}eh{\r u}{\v r}ek}, \bibinfo{author}{P.~Sojka},
\newblock \bibinfo{title}{{Software Framework for Topic Modelling with Large Corpora}},
\newblock in: \bibinfo{booktitle}{{Proceedings of the LREC 2010 Workshop on New Challenges for NLP Frameworks}}, \bibinfo{publisher}{ELRA}, \bibinfo{address}{Valletta, Malta}, \bibinfo{year}{2010}, pp. \bibinfo{pages}{45--50}. \bibinfo{note}{\url{http://is.muni.cz/publication/884893/en}}.
\bibitem[{Caselles-Dupr{\'e} et~al.(2018)Caselles-Dupr{\'e}, Lesaint, and Royo-Letelier}]{caselles2018word2vec}
\bibinfo{author}{H.~Caselles-Dupr{\'e}}, \bibinfo{author}{F.~Lesaint}, \bibinfo{author}{J.~Royo-Letelier},
\newblock \bibinfo{title}{Word2vec applied to recommendation: Hyperparameters matter},
\newblock in: \bibinfo{booktitle}{Proceedings of the 12th ACM Conference on Recommender Systems}, \bibinfo{year}{2018}, pp. \bibinfo{pages}{352--356}.
\bibitem[{Rauch et~al.(2024)Rauch, Schwinger, Wirth, Heinrich, Huseljic, Herde, Lange, Kahl, Sick, Tomforde et~al.}]{rauch2024birdset}
\bibinfo{author}{L.~Rauch}, \bibinfo{author}{R.~Schwinger}, \bibinfo{author}{M.~Wirth}, \bibinfo{author}{R.~Heinrich}, \bibinfo{author}{D.~Huseljic}, \bibinfo{author}{M.~Herde}, \bibinfo{author}{J.~Lange}, \bibinfo{author}{S.~Kahl}, \bibinfo{author}{B.~Sick}, \bibinfo{author}{S.~Tomforde}, et~al.,
\newblock \bibinfo{title}{Birdset: A large-scale dataset for audio classification in avian bioacoustics},
\newblock \bibinfo{journal}{arXiv preprint arXiv:2403.10380}  (\bibinfo{year}{2024}).
\bibitem[{Vaswani et~al.(2017)Vaswani, Shazeer, Parmar, Uszkoreit, Jones, Gomez, Kaiser, and Polosukhin}]{vaswani2017attention}
\bibinfo{author}{A.~Vaswani}, \bibinfo{author}{N.~Shazeer}, \bibinfo{author}{N.~Parmar}, \bibinfo{author}{J.~Uszkoreit}, \bibinfo{author}{L.~Jones}, \bibinfo{author}{A.~N. Gomez}, \bibinfo{author}{{\L}.~Kaiser}, \bibinfo{author}{I.~Polosukhin},
\newblock \bibinfo{title}{Attention is all you need},
\newblock \bibinfo{journal}{Advances in neural information processing systems} \bibinfo{volume}{30} (\bibinfo{year}{2017}).
\bibitem[{Tulkens and {van Dongen}(2024)}]{minishlab2024model2vec}
\bibinfo{author}{S.~Tulkens}, \bibinfo{author}{T.~{van Dongen}},
\newblock \bibinfo{title}{Model2vec: Fast state-of-the-art static embeddings}  (\bibinfo{year}{2024}). \URLprefix \url{https://github.com/MinishLab/model2vec}.
\bibitem[{PACE(2017)}]{PACE}
\bibinfo{author}{PACE}, \bibinfo{title}{{P}artnership for an {A}dvanced {C}omputing {E}nvironment ({PACE})}, \bibinfo{year}{2017}. \URLprefix \url{http://www.pace.gatech.edu}.

\end{thebibliography}
\appendix
\section{STSG v1 Hyperparameter Tuning}
\label{sec:stsg_v1}

The first version of STSG uses a spectrogram with 128 Mel bands and 20 MFCCs.

\begin{figure}[h!]
    \centering
    \includegraphics[width=0.6\columnwidth]{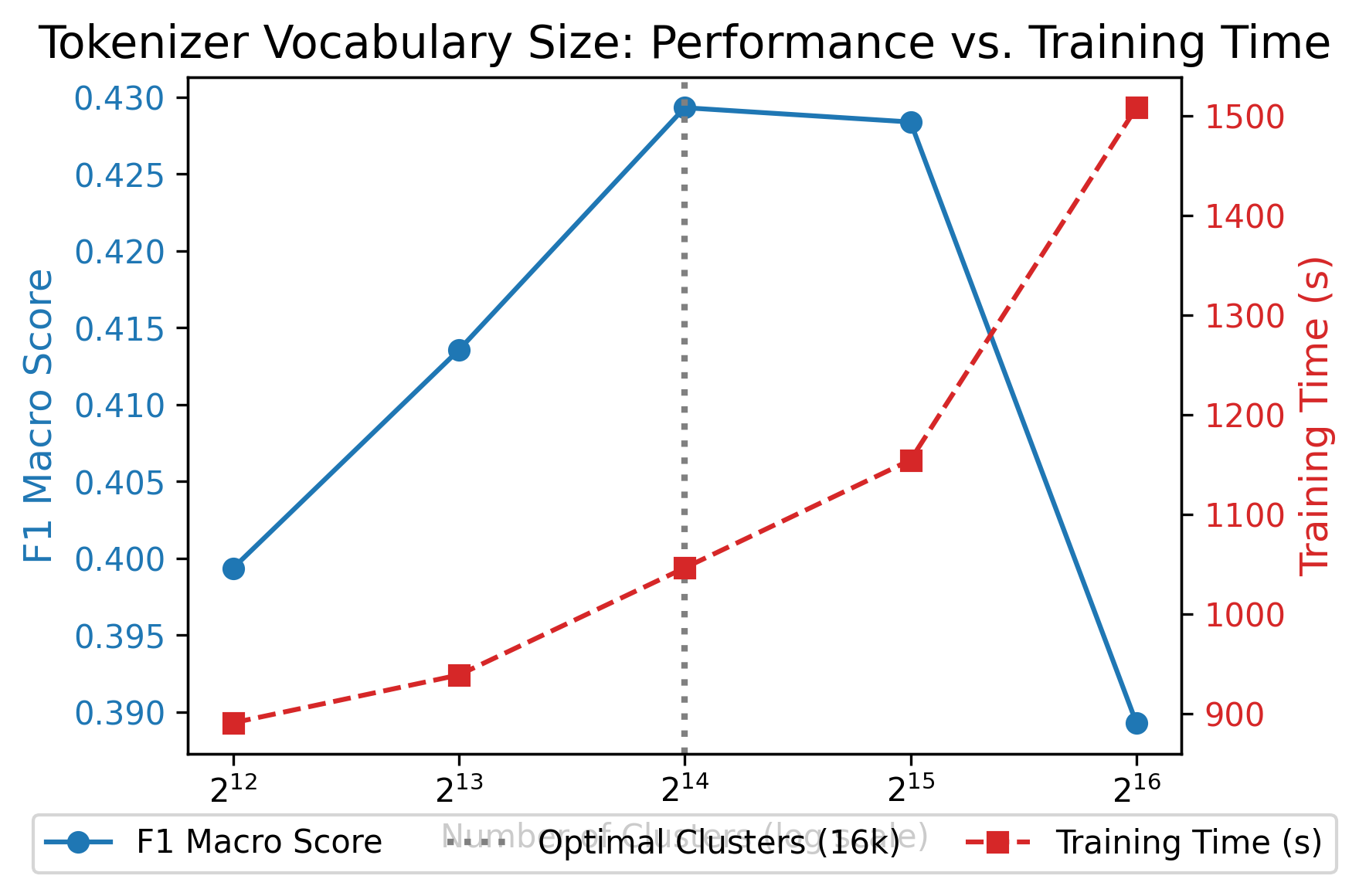}
    \caption{F1 score and training time vs. tokenizer vocabulary size. A vocabulary of 16k clusters provides the best performance before diminishing returns.}
    \label{fig:vocab_tradeoff}
\end{figure}

\begin{table*}[h!]
\centering
\caption{
 Word2Vec hyperparameter sweep results, including F1 Macro scores and training times on the surrogate validation task. 
 Deltas are calculated relative to the baseline configuration. 
 The best-performing row in each scan group is highlighted in bold.
}
\label{tab:w2v_full_sweep}
\begin{tabular}{llccrrc}
\toprule
\textbf{Scan Group} & \textbf{Parameter} & \textbf{Value} & \textbf{F1 Macro} & \textbf{$\Delta$ F1} & \textbf{Time (s)} & \textbf{$\Delta$ Time (s)} \\
\midrule
\multirow{1}{*}{Baseline} 
 & \texttt{vector\_size} & 256 & 0.424 & +0.000 & 1108 & +0 \\
 & \texttt{window} & 80 & & & & \\
 & \texttt{ns\_exponent} & 0.75 & & & & \\
 & \texttt{sample} & \textit{1e-4} & & & & \\
\midrule
\multirow{4}{*}{Varying \texttt{vector\_size}} 
 & \texttt{vector\_size} & 128 & 0.381 & -0.043 & 934 & -174 \\
 & \texttt{vector\_size} & 384 & 0.449 & +0.025 & 1199 & +91 \\
 & \texttt{vector\_size} & 512 & 0.445 & +0.021 & 1674 & +566 \\
 & \texttt{vector\_size} & 1028 & \textbf{0.495} & \textbf{+0.071} & \textbf{2828} & \textbf{+1720} \\
\midrule
\multirow{2}{*}{Varying \texttt{window}} 
 & \texttt{window} & 40 & 0.426 & +0.002 & 570 & -538 \\
 & \texttt{window} & 120 & \textbf{0.427} & \textbf{+0.003} & \textbf{1480} & \textbf{+372} \\
\midrule
\multirow{2}{*}{Varying \texttt{ns\_exponent}} 
 & \texttt{ns\_exponent} & 0.0 & \textbf{0.440} & \textbf{+0.016} & \textbf{1050} & \textbf{-58} \\
 & \texttt{ns\_exponent} & -0.5 & 0.401 & -0.023 & 1043 & -65 \\
\midrule
\multirow{2}{*}{Varying \texttt{sample}}
 & \texttt{sample} & \textit{1e-5} & \textbf{0.423} & \textbf{-0.001} & \textbf{589} & \textbf{-519} \\
 & \texttt{sample} & \textit{1e-6} & 0.379 & -0.045 & 121 & -987 \\
\midrule
\multirow{1}{*}{Combination} 
 & \texttt{window} & 120 & 0.405 & -0.019 & 1471 & +363 \\
 & \texttt{ns\_exponent} & -0.5 & & & & \\
\bottomrule
\end{tabular}
\end{table*}

\begin{figure}[h!]
    \centering
    \begin{subfigure}[b]{0.48\columnwidth}
        \includegraphics[width=\textwidth]{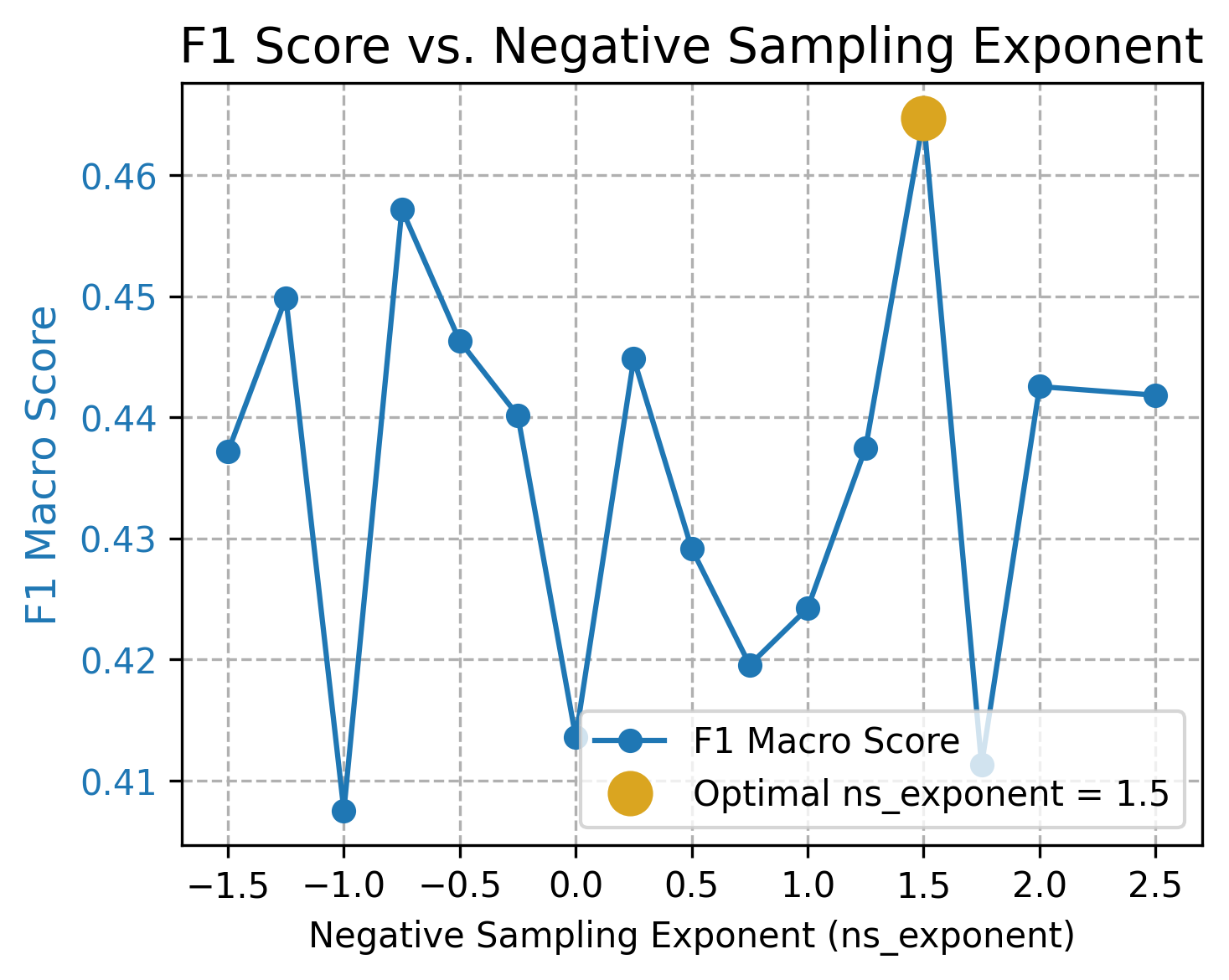}
        \caption{F1 score vs. ns\_exponent.}
        \label{fig:ns_exponent_tune}
    \end{subfigure}
    \hfill 
    \begin{subfigure}[b]{0.48\columnwidth}
        \includegraphics[width=\textwidth]{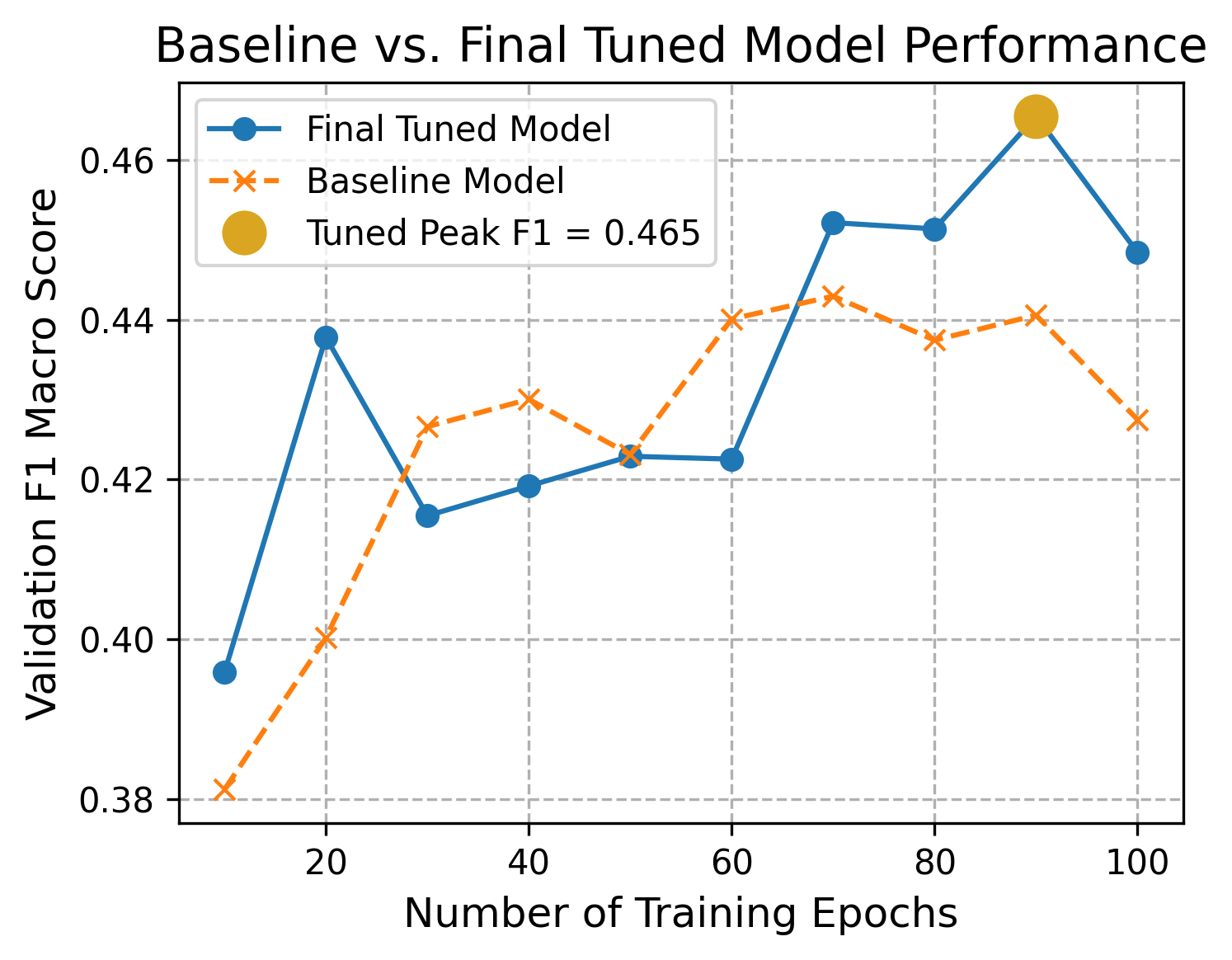}
        \caption{Baseline vs. Tuned Model.}
        \label{fig:comparison_curve}
    \end{subfigure}
    \caption{Hyperparameter tuning and final model performance. (a) Tuning the negative sampling exponent shows a clear peak at 1.5. (b) The final tuned model consistently outperforms the baseline.}
    \label{fig:tuning_and_comparison}
\end{figure}
\end{document}